\documentclass[iop,appendixfloats]{emulateapj}
\usepackage{tablefootnote}
\usepackage{appendix}
\usepackage{amsmath}
\usepackage{amssymb}
\usepackage{natbib}
\usepackage{mathtools}
\usepackage{graphicx}

\newenvironment{frcseries}{\fontfamily{frc}\selectfont}{}

\newcommand{\vect}[1]{\boldsymbol{\mathbf{#1}}}

\newcommand{\eqn}[1]{\begin{align*}
#1
\end{align*}}

\newcommand{\Bl}{\Big\{}
\newcommand{\Br}{\Big\}}

\newcommand{\mtx}[4]{
\[
#1 = #2
\left[ {\begin{array}{#3}
 #4
 \end{array} } \right]
\]
}

\newcommand{\eqnset}[4]{
\[ #1 = #2 \left\{ \begin{array}{#3}
        #4
\end{array} \right. \] 
}

\newcommand{\vx}{\vect{x}}
\newcommand{\vy}{\vect{y}}

\newcommand{\hesslld}[2]{\frac{\partial^2 \llpp}{\partial A_{#1} \partial A_{#2}}}
\newcommand{\sumn}{\sum^n_{i=1}}
\newcommand{\summ}{\sum^m_{j=1}}
\newcommand{\summo}{\sum^{m-1}_{j=1}}

\newcommand{\llpp}{\ell({\bf A}_{EM}^{\prime})}

\def\tilde{\widetilde}

\def\beq{\begin{equation} }
\def\eeq{\end{equation} }
\def\spose#1{\hbox to 0pt{#1\hss}}
\def\ltsim{\mathrel{\spose{\lower.5ex\hbox{$\mathchar"218$}}
     \raise.4ex\hbox{$\mathchar"13C$}}}

\def\tilde{\widetilde}
\def\spose#1{\hbox to 0pt{#1\hss}}
\def\lta{\mathrel{\spose{\lower 3pt\hbox{$\mathchar"218$}}
        \raise 2.0pt\hbox{$\mathchar"13C$}}}
\def\gta{\mathrel{\spose{\lower 3pt\hbox{$\mathchar"218$}}
        \raise 2.0pt\hbox{$\mathchar"13E$}}}

\begin{document}

\title{Reconstructing the Accretion History of the Galactic Stellar Halo from chemical abundance ratio distributions}
\shorttitle{Accretion History of the Galactic Halo.}
\shortauthors{Lee et al.}

\author{Duane M. Lee\altaffilmark{1}, Kathryn V. Johnston\altaffilmark{2}, Bodhisattva Sen\altaffilmark{3}, Will Jessop\altaffilmark{3}}
\altaffiltext{1}{Research Center for Galaxies and Cosmology, Shanghai Astronomical Observatory, Chinese Academy of Sciences, 80 Nandan Road, Shanghai, 200030, China; duane@shao.ac.cn}
\altaffiltext{2}{Department of Astronomy, Columbia University, New York City, NY 10027}
\altaffiltext{3}{Department of Statistics, Columbia University, New York City, NY 10027}

\begin{abstract}
Observational studies of halo stars during the last two decades have placed some limits on the quantity and nature of accreted dwarf galaxy contributions to the Milky Way stellar halo by typically utilizing stellar phase-space information to identify the most recent halo accretion events. In this study we tested the prospects of using 2-D chemical abundance ratio distributions (CARDs) found in stars of the stellar halo to determine its formation history. First, we used simulated data from eleven ``MW-like'' halos to generate satellite template sets of 2-D CARDs of accreted dwarf satellites which are comprised of accreted dwarfs from various mass regimes and epochs of accretion. Next, we randomly drew samples of $\sim10^{3-4}$ mock observations of stellar chemical abundance ratios ([$\alpha$/Fe], [Fe/H]) from those eleven halos to generate samples of the underlying densities for our CARDs to be compared to our templates in our analysis. Finally, we used the expectation-maximization algorithm to derive accretion histories in relation to the satellite template set (STS) used and the sample size. For certain STS used we typically can identify the relative mass contributions of all accreted satellites to within a factor of 2. We also find that this method is particularly sensitive to older accretion events involving low-luminous dwarfs e.g. ultra-faint dwarfs --- precisely those events that are too ancient to be seen by phase-space studies of stars and too faint to be seen by high-z studies of the early Universe. Since our results only exploit two chemical dimensions and near-future surveys promise to provide $\sim6-9$ dimensions, we conclude that these new high-resolution spectroscopic surveys of the stellar halo will allow us to recover its accretion history --- and the luminosity function of infalling dwarf galaxies --- across cosmic time. 
\end{abstract}

\keywords{Galaxy: abundances --- Galaxy: halo --- Galaxy: stellar content --- galaxies: dwarf --- galaxies: early universe --- stars: abundances}

\maketitle


\section{Introduction}\label{ahp_sec:intro}
The origin of the stellar halo has been a topic of intense study since the publication of the seminal paper by \cite{eggen62}. The paper suggested that the stellar halo originated from the ``monolithic collapse'' of a galactic-sized primordial gas cloud. More specifically, they proposed that during this quick ($\lesssim100$ Myrs) collapse a very small portion of that metal-poor/free gas fragmented, due to Jeans' instabilities, and formed stars. While the bulk of the gas would eventually form the young, metal-rich, circularly-orbiting, stellar disk of the Galaxy, these ``halo'' stars would instead be characterized as old, metal-poor, stars on mainly radial orbits due to the imprint of the cloud's initial collapse. When \cite{eggen62} proposed this theory observations of the halo were restricted to small kinematic samples near the Sun --- samples which lacked any features that might suggest that the halo was built over time via galactic mergers or accretion.
However, a decade and a half later, \cite{searle78} stated in another seminal work that, in fact, some halo observations could be explained in another way. Their paper advanced the idea that differences in globular cluster abundance distributions versus galacto-centric distances in the halo were due to the ``hierarchical merging'' of many smaller galactic systems over the lifetime of the Galaxy. As a consequence of hierarchical merging, the stellar halo was created metal-poor because most galactic progenitors of the halo were accreted early on, which, in turn, afforded stellar inhabitants of these accreted systems little time to enrich to higher metallicities. The theory also suggested that, while less abundant, a distribution of more metal enriched stars and clusters should also inhabit the halo due to mergers over time. Consequently, it was these mergers that led to the radial orbits of stars and clusters that were earlier seen and characterized by \cite{eggen62}. 

Also bolstering the theory of hierarchical merging was the development of the theories of the formation of structures within the cold dark matter paradigm \citep[e.g.,][]{efstathiou85}. These theories predicted that the continuous merging of galaxies was facilitated by the parallel growth of the dark matter halos that hosted or formed the backbones of those galaxies.  As a consequence, hierarchical merger formation of the stellar halo is simply a manifestation of that growth at the galactic scale.

While cosmological theory supported Searle \& Zinn's work, strong additional evidence for the theory of hierarchical merging came with the observations of halo substructure. In the early 90's, \cite{ibata94} discovered the core of the Sagittarius dwarf galaxy in the outskirts of the stellar halo. Observations of this obviously ``dying'' satellite supported the assertion that stellar debris from the dwarf would follow the orbit of the accreted system. This debris would also disperse in phase-space over time and contribute to the growth of the halo. Further evidence for hierarchical merging came from the Sloan Digital Sky Survey \citep[SDSS;][]{york00}. This state-of-the-art project was the first global survey of the halo to extend beyond a couple of kiloparsecs from the Sun. All previous deep surveys of the halo were done in pencil beam mode --- a mode where missing extended structures was virtually guaranteed. Initial results from SDSS showed a halo teeming with photometric overdensities within $\sim$18 kpc from the galactic center. This finding suggested that substructure was ubiquitous \citep{newberg02}. \cite{majewski03} found the tidal tails of Sagittarius wrapped around the MW by observing M-giant overdensities in the halo. The ``smoking gun'' for hierarchical merging came in 2006, when a clear and distinct photometric picture of the halo from SDSS revealed newly discovered dwarf galaxies and, more to the point, tidal streams (i.e. substructure) from past mergers called the ``field of streams'' \citep{belokurov07a}.

The SDSS discoveries of abundant substructure in the halo led to numerous dynamical studies. Some studies determined the membership of known objects \citep[e.g][]{majewski05} while others discovered new objects by their dynamical overdensities in phase-space \citep[e.g.][]{schlaufman09}. Beyond SDSS lies the next generation of galactic halo surveys. From photometry (LSST), astrometry (Gaia), and high-res abundances (APOGEE \& GaLAH), we can expect to collect enough data for use in statistical analysis to actually answer some of the outstanding questions in Galactic astronomy. One outstanding question of great importance is: what is the merger history of the MW halo? With the aforementioned surveys soon at our disposal, we will have three ways of approaching this question.

A traditional photometric census of the halo (LSST) is only sensitive to mergers that are a few billion years old due to the phase-mixing of the projected phase-space dimensions of accreted structures \citep{sharma10}. Dynamical studies like Gaia should prove more successful in recovering accretion histories because these studies collect data that contains full 6-D phase-space information. In fact, in principle, this information allows one to calculate orbital properties (i.e., integrals of motion) for a given potential. Since the integrals of motion for a static potential are conserved, it is possible to associate debris in orbital-property space even if the halo is fully phase-mixed \citep{helmi00}. However, for the outer halo (beyond 10 kpc), even Gaia cannot measure distances with sufficient accuracy, and this means that reconstructed histories of this depth (via astrometric data) are still incomplete. Furthermore, it is highly likely that rapidly occurring, violent mergers took place in the early assembly of the halo. Significant mergers of this nature should scatter normally-conserved quantities in phase-space making the extraction of merger histories from earlier epochs harder, and perhaps futile. 

In the past decade, an understanding of the limitations to stellar phase-space data analysis has led of the promising pursuit of conserved quantities in stellar chemical abundance space --- that is, stellar quantities which are more innate and, as such, cannot be changed by scattering in phase-space. \cite{unavane96} were the first to demonstrate that such innate quantities could be fruitful by using a metallicity-color ([Fe/H]-($B$-$V$)) plane to select halo stars, which are similar in composition to existing metal-poor dSph satellite stars, to constrain the hierarchal buildup of the halo. Using this comparison, Unavane determined that the history of the halo cannot contain more than $\sim60$ Carina-like dwarf accretions or $\simeq6$ Fornax-like dwarf accretions. In an analogous proposal for the Galactic disk, Freeman \& Bland-Hawthorn \citep{freeman02,bland-hawthorn04} suggested that measuring the detailed chemical composition of vast numbers of the stars in the Galactic disk might be used to recover their origins: those with identical compositions in high-dimensional abundance space are likely to have been born in the same star cluster. \cite{de-silva07} observed that star clusters are chemically homogenous within error while \cite{bland-hawthorn10} confirmed that this homogeneity allows astronomers to track stars back to the natal clusters by ``chemically tagging'' these stars. Thus ``chemical tagging" could be used to reconstruct long-dead star clusters and recover the SFH of the Galaxy.  

\begin{figure}[htp]
\begin{center}
   \includegraphics[width=0.475\textwidth,angle=0]{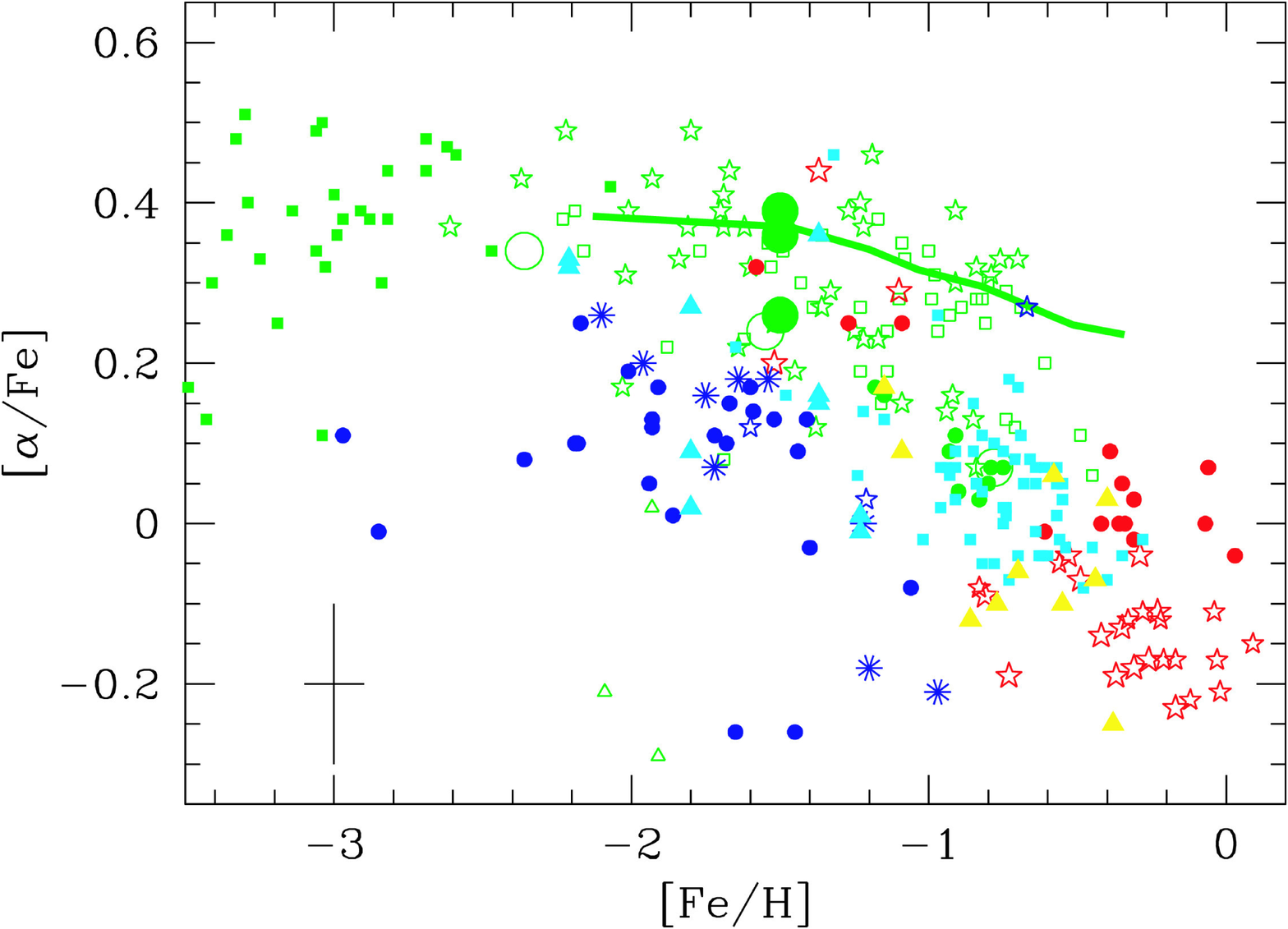}
\caption{This figure is a reproduction of Figure 12 from \cite{geisler07}. The figure is a compilation of [$\alpha$/Fe] vs. [Fe/H] data taken by \cite{nissen97,ivans99,shetrone01,venn01,fulbright02,smecker-hane02,stephens02,gratton03,shetrone03,venn03,bonifacio04,cayrel04,kaufer04,geisler05,jonsell05,monaco05,johnson06,pompeia06,tautvaisiene07}. Symbols shown here represent a mixture of model data, stars and star clusters found in the MW halo (green), as well as stars and stellar clusters found in low-mass dwarf spheroidals (blue), dwarf irregulars (yellow), the Sagittarius dwarf galaxy (red), and the large Magellanic Cloud (cyan). The distribution of accreted and ``soon-to-accreted'' systems in this 2-D chemical abundance space demonstrates the potential for determining accretion histories by attributing various subsets of the chemical abundance ratio distributions (CARDs) observed in the stellar halo of a nearby galaxy (e.g., the MW halo) to different accreted systems (see text for brief explanation).}
\label{ahp_fig:Geisler}
\end{center}
\end{figure}

In this paper we explore whether a modified version of "chemical tagging" might be applied to the Galactic halo, expanding on the idea that \cite{unavane96} introduced over a decade earlier.  Unlike stars born in the same cluster, stars born in the same dwarf galaxy do NOT share the same chemical composition. However, pioneering studies in the last decade have shown that stars in different dwarfs do have distinct (if overlapping) chemical abundance ratio distributions \citep[CARDs; see, e.g.,][]{nissen97,ivans99,shetrone01,venn01,fulbright02,smecker-hane02,stephens02,gratton03,shetrone03,venn03,bonifacio04,cayrel04,kaufer04,geisler05,jonsell05,monaco05,johnson06,pompeia06,tautvaisiene07}. Figure~\ref{ahp_fig:Geisler} from \cite{geisler07} illustrates how these CARDs (revealed from a compilation of the aforementioned observations) tantalizingly suggests that such an attempt is possible. 

Figure~\ref{ahp_fig:Geisler} is a reproduction of Figure 12 in \cite{geisler07} showing a 2-D CARDs plot of the [$\alpha$/Fe] (the ratio of the sum of $\alpha$-elements (typically, Ca, Mg, Ti, \& O) to Fe) versus [Fe/H]. The plot shows various different star and star cluster measurements of [$\alpha$/Fe] and [Fe/H] which separate different parent or host systems into different parts of the 2-D CARD space. Additionally, differences between different galactic systems at lower metallicities are also emerging for neutron-capture elements (e.g., Strontium and Barium). These observations suggest that 

\begin{itemize}
\item at a given accretion epoch, differences (in CARDs) between systems of the different stellar masses exist
\item at a given stellar mass, differences between systems that were accreted at different times exist
\end{itemize}

In this paper, we develop a statistical approach (that uses the EM algorithm) to examine whether the CARDs of different mass objects accreted at different times are sufficiently different to allow us to recover halo accretion histories using data alone. We test our method with the semi-analytic models available from previous simulation work. In \S\ref{ahp_sec:Meth}, we explain the nature of the models and methods used to produce accounts of accretion history from mock halo observations. In \S\ref{ahp_sec:results1}, we discuss the success of the EM algorithm when applied to specific cases. In \S\ref{ahp_sec:results2}, we describe the success of our results across our entire set of data. In \S\ref{ahp_sec:diss}, we discuss both the utility and reliability of applying this technique to real observations. In \S\ref{ahp_sec:conc}, we present our conclusions.

\section{Methods}\label{ahp_sec:Meth}
We can approach the problem of recovering the accretion history of a galactic halo (using CARDs) by posing the following question: 
\begin{quote}
``How accurately can we determine the fraction of total stellar mass, $A_{j}$, contributed by satellites of various mass ($M_{sat}$) and accretion time ($t_{acc}$) to a stellar halo given a set of templates for the distribution $f_{j} (x_{d},M_{sat},t_{acc})$ of chemical abundances $x_{d}$ found in those satellites, and observations of CARDs ($f$($x_{d}$)) in the stellar halo?'' 
\end{quote}
In this study, we attempt to answer this question by investigating realizations of the stellar halo by Bullock \& Johnston (2005; see \S\ref{ahp_sec:Sims}) which includes distributions of $\alpha$- and iron (Fe) elements generated by the methods of \cite{robertson05} and implemented in the models by \cite{font06}. To begin our investigation, we define our approach by recasting our question in the form of the following equation: 
\begin{equation}
    f (x_{d}) = \sum_{j}^{m}{A_{j}\cdot f_{j}  (x_{d},M_{sat},t_{acc})}\label{ahp_eq:mixture}
\end{equation}
where $$\sum_{j}^{m}{A_{j}} = 1$$ for $m$ satellite templates with $A_{j}\ge0$.

In Eqn.~\ref{ahp_eq:mixture}, $f$($x_{d}$) represents the probability density function (distribution) of observed ``stars'' in the $d$-dimensional CARD space ($x_{1,2,3, ... d}$) and $A_{j}$ represents the relative contributions from each template $f_{j}$. In a generic sense, each template $f_{j}$ represents the CARD for dwarfs of some characteristic mass $M_{sat}$ that were accreted at a characteristic time $t_{acc}$. Hence, finding all $A_{j}$ values corresponds to recovering the ``accretion history profile'' (AHP) of the galactic halo. Utilizing Eqn.~\ref{ahp_eq:mixture} to address our question requires the following four steps:

\begin{enumerate}
\item Generate mock ``observations'' of CARDs (i.e. $f$($x_{d}$) in our case with [$x_{1},x_{2}$] = [[$\alpha$/Fe],[Fe/H]]) for 11 realizations from simulations of purely accretion-grown halos (\S\ref{ahp_sec:Obs}).
\item Create CARD templates ($f_{j} (x_{d},M_{sat},t_{acc})$) representing the density of stars in [$\alpha$/Fe]-[Fe/H] space for satellites found in selected 2-D bins of satellite mass and accretion time (\S\ref{ahp_sec:STS}).
\item Apply the expectation-maximization (EM) algorithm (a method for statistical estimation in finite mixture models [see \S\ref{ahp_sec:EM}]) to observations using satellite templates to recover their relative contribution (i.e. $A_{j}$) to the host halo's stellar mass (\S\ref{ahp_sec:results1} and \S\ref{ahp_sec:results2}).
\item Evaluate the efficacy of this approach by using a summary statistic (\S\ref{ahp_sec:ATFoE}) to encapsulate how accurate the method is in recovering the known accretion histories for each halo (e.g., see \S\ref{ahp_sec:all}).
\end{enumerate}
 
\subsection{The Simulations}\label{ahp_sec:Sims}
The simulations consist of 11 ``MW-sized'' halo realizations which involve a total of 1515 accreted satellites (with $100-150$ satellites contributing to each halo) from the \cite{bullock05} work. Each dark matter host of the 11 halo realizations has a total mass of $M_{virial}$($z=0$) = 1.4 x 10$^{12} M_{\odot}$ generated by merger trees using a statistical Monte Carlo method with an extended Press-Schecter (EPS) formalism \citep[][and references therein]{somerville99,lacey93,bullock05}. Differences in the AHP between each halo are entirely based on the randomness in the merger trees. 

CARDs for these 11 merger histories were generated from a semi-analytic chemical enrichment code \cite{robertson05} which was applied separately to each infalling satellite and combined with the simulations by \cite{font06}.  Since the enrichment code was implemented for each satellite generated, we can look at individual satellites to assess their relative contribution to their host halo's CARDs. Also, since the aim of this study is to determine the amount of information we can retrieve via chemical abundance observations, we abstain from utilizing any of the satellites' spatial information in our analysis. The main factors contributing to the the star formation history in the satellites are (1) the epoch of reionization, $z_{re}$, (2) the fraction of gas remaining/accreted in the satellite halo after reionization (set mainly by the satellite's virial mass at its time of accretion), (3) the global star formation rate, and (4) the termination of star formation at the time of accretion \citep{bullock05}. These  parameters are utilized in the simulations to determine the amount of gas available to produce stars and the duration of star formation, which, in turn, determines the chemical evolution of each satellite as prescribed in \cite{robertson05}. The prescription includes $\alpha$- and Fe-element enrichment from Type II and Type Ia supernovas and stellar wind outflows of metals. The chemical evolution model was tuned with a supernova (SN) feedback treatment to agree with the local dwarf galaxy stellar mass-metallicity \citep[][; see \S\ref{ahp_sec:STS} for further discussion]{robertson05}. The $\alpha$-element patterns in dwarfs versus the smooth halo are consistent with the CARDs of dwarfs found in the compilation of data in Figure 12 of \cite{geisler07} (see Figure~\ref{ahp_fig:Geisler}) --- an agreement that further bolsters our approach in this investigation \citep{font06}.

\subsection{``Observations" from the Simulations}\label{ahp_sec:Obs}
The function $f$(x$_{d}$) represents the density distribution produced by $n$ random ``observations'' in chemical abundance space $x_{d}$ of ``stars'' (star particles; see \S\ref{ahp_sec:Sims} for explanation). Sample distributions for each halo are constructed by randomly drawing ``stars''.  To mimic observational errors during mock observations, we add a random number drawn from a Guassian with a dispersion of 0.05 dex to both [$\alpha$/Fe] and [Fe/H] abundance ratios. The choice of the size of these errors is meant to probe the foreseeable potential of this technique by employing the best possible conditions for analysis. Evaluation of this technique with ideal conditions provides us with a baseline for expectations from which analysis of real observations in the future can be assessed. In our study, we select samples of ${nd} \approx 10^{3}$, $10^{4}$, and $3\times10^{4}$ representing current, near-future, and optimistically-anticipated sample sizes, respectively (Ken Freeman, private communication). 

\begin{figure}[t]
   \centering
  \includegraphics[width=0.475\textwidth,angle=0]{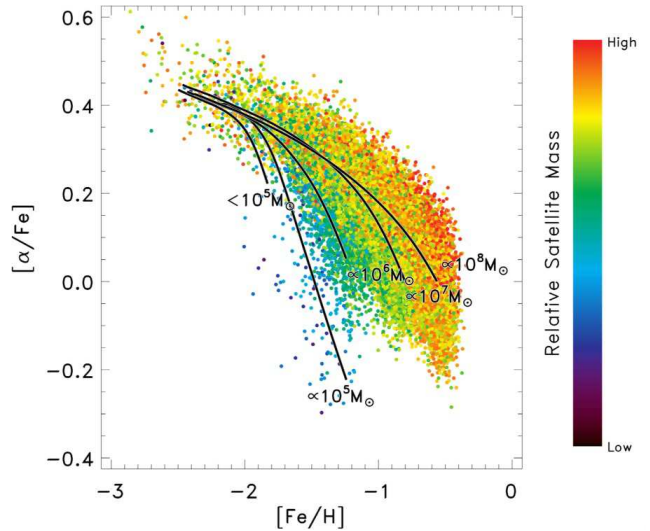} 
   \caption{Plot of [$\alpha$/Fe] vs. [Fe/H] for $\sim3\times10^{4}$ ``star particles.'' Each particle is color-coded to represent the relative stellar mass/luminosity of its parent satellite. The relative number of particles in the accreted satellite mass/luminosity range reflects the expected relative contribution from each parent to the total stellar mass of the host halo. The chemical evolution tracks of five satellites, randomly chosen to span the stellar mass range of accreted satellites for halo 1, are displayed over the colored particle distribution as black lines and labeled by a stellar mass proportional to the typical satellite stellar mass found in the mass bins outlined in \S\ref{ahp_sec:STS} and displayed in Figure~\ref{ahp_fig:5x5}.}
   \label{ahp_fig:2-D_cad}
\end{figure}

Figure~\ref{ahp_fig:2-D_cad} shows a 2-D CARD ([$\alpha$/Fe] vs. [Fe/H]) of $\sim3\times10^{4}$ star particles representing mock stellar abundance ratio observations from the halo 1 simulation. The color of each particle represents the stellar mass of its parent satellite relative to all other accreted satellites. Black and purple particles are donated from the least massive satellites while orange and red particles are donated from the most massive satellites. The distribution of particles shown demonstrate the expectation that the most massive satellites should account for the vast majority of stars found in the accreted halo stellar population. In comparing this 2-D CARD to the observed CARDs in Figure~\ref{ahp_fig:Geisler}, we see that the distributional spread between observed accreted dwarfs of different masses mirror the distributional spread (in mass) for the simulated dwarfs. 

The black dashed lines that overlay the colored particle distribution of Figure~\ref{ahp_fig:2-D_cad} represent chemical evolution tracks (from the simulations) of typical dwarf masses accreted over the lifetime of the halo. The length of these tracks are primarily affected by the satellite's accretion time. The more time a satellite has to produce stars, the longer its galactic chemical evolution can continue to advance to higher metallicities, and vice versa. The curvature of these tracks is primarily determined by the satellite's mass. The more mass a satellite has to produce stars, the higher its star formation rate (SFR), which means chemical enrichment by core-collapse SN is greater. This enhanced early enrichment from core-collapse SN leads to higher galactic metallicities before the typical 1 Gyr onset (delay) in Type Ia SN begins (ends) to establish a so-called [$\alpha$/Fe]-knee via significant contributions to Fe abundances. The incorporation of these various tracks into our dwarf model templates are discussed in the next section.
\subsection{Satellite Template Sets}\label{ahp_sec:STS}
\begin{figure*}[pt]
\vspace{.45in}
\vspace{-.45in}
\begin{center}
\begin{tabular}{c}
\begin{tabular}{l}
\vspace{.145in}
  \includegraphics[height=3.951in,angle=0]{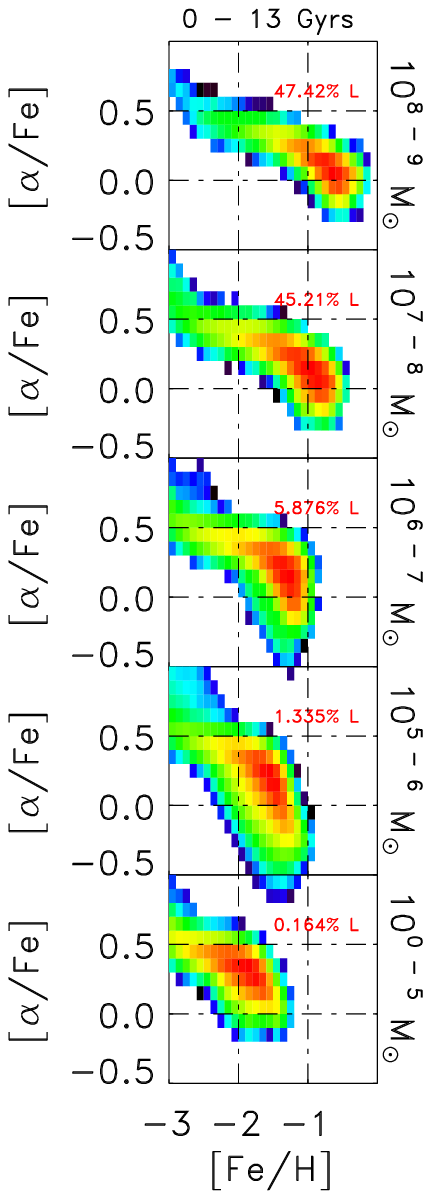}\\
\vspace{-.1in}\hspace{-.1in}\\
 \includegraphics[width=1.434in,angle=0]{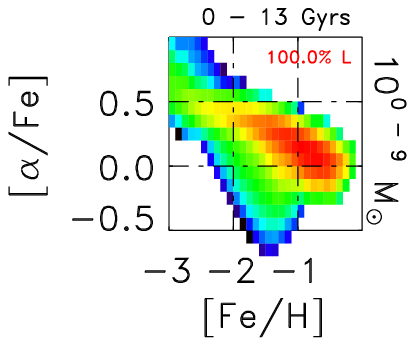}   
 \end{tabular}
 \hfill
 \begin{tabular}{l}
  \includegraphics[width=4.5in,angle=0]{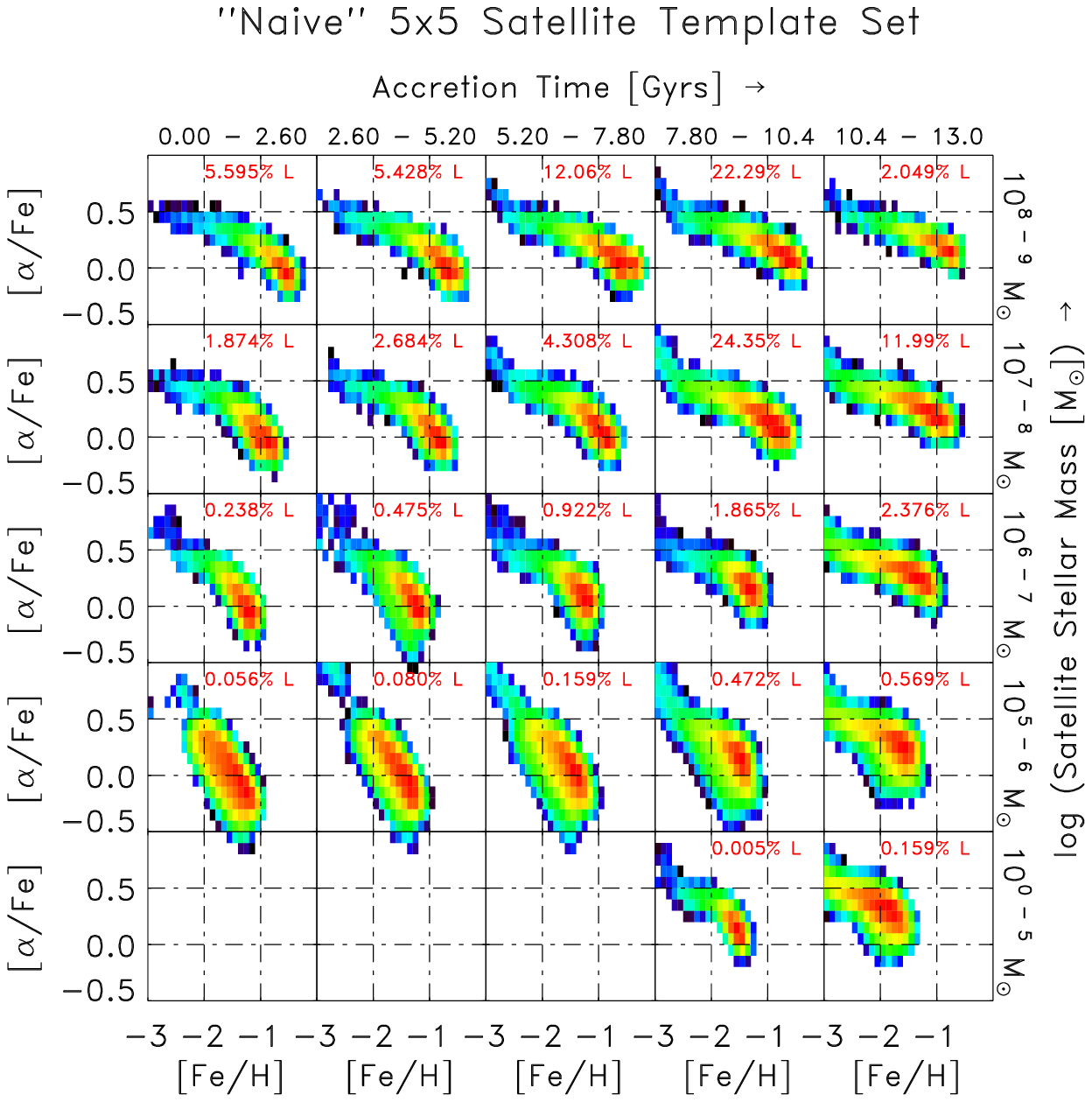}\\
\vspace{0in}\\
    \includegraphics[width=4.3in,angle=0]{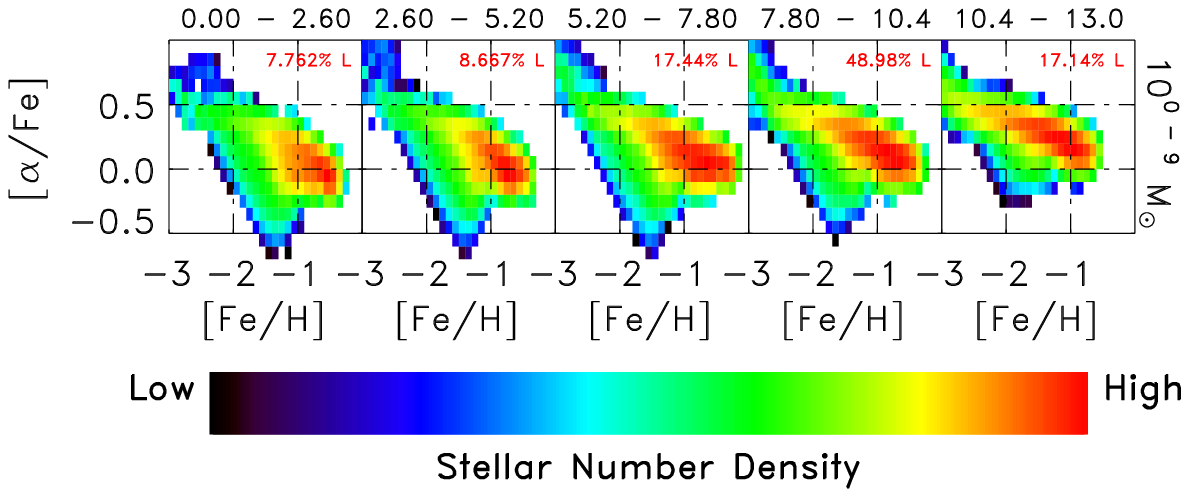}
\end{tabular}
\end{tabular}  
   \caption{Plot of 5x5 STS along with projects in the $t_{acc}-M_{sat}$ plane. {\it Top-right}: Our 5x5 STS. The relative contribution of stellar mass from a subset of all 1515 satellites in each template is shown as percentages of the total halo stellar mass (red). Each column and row reflects the mass/stellar mass-metallicity relation and age-metallicity relation, respectively (see \S\ref{ahp_sec:Sims} for details). {\it Top-left and bottom-right}: Projections of the 5x5 STS into the $t_{acc}$ plane ({top-left}) and $M_{sat}$ plane ({bottom-right}) are equivalent to the 1x5 (mass-divided) STS and 5x1 (time-divided) STS explored in \S\ref{ahp_sec:results1}. 
{\it Bottom-left}: Plot of a projection into both parameter dimensions exemplifies a density distribution (i.e. F($x_{d}$)) similar to the parent distributions of individual halos from which ``observed'' stars are drawn in our analysis.}\label{ahp_fig:5x5}
\end{center}     
\end{figure*}

To see if we can recover the AHP of our simulated halos from our mock observations we need to generate templates which represent typical accretion events of given satellite stellar mass and age. The most ``naive" approach to creating our templates would be to evenly divide the possible range in time $t_{acc}$ ($0-13$ Gyrs) and mass (stellar) $M_{sat}$ ($10^{0-9} M_{\odot}$). This division would form $N_{r}$ mass-binned templates (rows) by $N_{c}$ time-binned templates (columns) with some ``empty'' templates ($N_{empty}$) where the total number of templates equal $N_{temps} = N_{r}\times N_{c} - N_{empty}$. However, since decades in galactic (stellar) mass have intuitive implications for galaxy evolution, we restrict our current templates to even divisions in $t_{acc}$ while we divide $M_{sat}$ by decades of mass from $10^{5} M_{\odot}$ to $10^{9} M_{\odot}$ and combine all satellites below $10^{5} M_{\odot}$ into a $5^{th}$ mass bin (see Figure~\ref{ahp_fig:5x5}). 

After divisions in the $t_{acc}$ -- $M_{sat}$ plane are selected, all 1515 dwarf satellite models are divided amongst the bins created by the selected partitions based on each dwarf's individual $t_{acc}$ and $M_{sat}$. During the process, each dwarf's chemical track (see \S\ref{ahp_sec:Obs}) is smeared out by a convolution of each star particle with an observational error of $\sigma_{err}$ = 0.05 dex in both chemical dimensions. To generate the CARDs required for implementation of our recovery algorithm (i.e. the EM algorithm), we separate an average of $\sim19,500$ star particles per satellite (with errors) into square bins of 0.1 dex that span 3 dex in [Fe/H] (-3 -- 0 dex) and 1.7 dex in [$\alpha$/Fe] (-0.7 -- 1). The collection of all binned distributions in our 2-D chemical space are normalized to produce an ensemble of probability densities that represent our satellite template set (STS). 

Figure~\ref{ahp_fig:5x5} shows our 5x5 STS as an example of our model template scheme. The full 5x5 panel (top-right) shows the evenly-spaced bins in accretion time versus bins spaced out by decades of accreted satellite stellar mass down to $10^{5} M_{\odot}$, below which all other satellites are binned together. As stated in \S\ref{ahp_sec:Sims}, the feedback prescriptions in the chemical evolution models were tuned to reproduce the chemical abundance relationships observed in galactic surveys. First, the mass (luminosity) versus metallicity ([Fe/H]) relationship can be seen by inspecting the trends along any accretion time column. This relationship shows an increase in the distribution peak value of [Fe/H] (and a decrease in the distribution peak value of [$\alpha$/Fe]) with increasing mass (luminosity) of the galaxy. Second, an  age-metallicity relationship can be seen by inspecting the trends along any accreted satellite mass row (i.e. when holding the mass range constant). This relationship shows an decrease in the distribution peak value of [Fe/H] (and a increase in the distribution peak value of [$\alpha$/Fe]) with an increase in the accretion time epoch (i.e. which dictates the available time for star formation) of the galaxy. Although, it should be noted that this age-metallicity relationship is not strictly expected to hold for any given set of dwarf galaxies as other processes are as likely to quench star formation in dwarfs before accretion takes place.

Projections of the 5x5 STS, in accreted satellite mass and accretion time, are shown in top-left and bottom-right corners of Figure~\ref{ahp_fig:5x5}, respectively. A comparison of both projections reveals smaller differences in CARDs between adjacent bins of accretion time than in adjacent bins of accreted satellite mass. The similarities between dwarf models in the 5x1 STS projection suggests that the EM algorithm will perform better when utilizing the 1x5 STS projections of accreted satellite mass for estimates (see \S\ref{ahp_sec:ATF}, \S\ref{ahp_sec:sum_FoE} \& \S\ref{ahp_sec:all} for further discussion). Finally, a 1x1 STS projection displaying the probability density function of our master template (i.e. containing the CARDs of all 1515 simulated dwarfs) is shown in the bottom-left of the figure.

In \S\ref{ahp_sec:results1}, we use the two 1-D projections discussed here to form a basis of analysis for the EM algorithm's performance and our ability to recover AHP of halos in one dimension of mass or time.

\subsection{Recovering AHPs using the EM Algorithm}\label{ahp_sec:EM}
The composition of our halos can be best described as a finite mixture of discrete accreted objects that exhibit varying characteristics in a shared CARD space ($x$=[$\alpha$/Fe],$y$=[Fe/H]). Since we can construct models for these accreted objects, we can create a mixture model
\begin{equation}\label{ahp_eq:mixmod}
f(x_{i},y_{i}) = \sum_{j=1}^{m}A_{j}\,f_{j}(x_{i},y_{i})
\end{equation}
where the relations $$\sum_{j=1}^{m}A_{j} = 1\;\;\;\;{\rm for}\; A_{j} \geq 0,	\;\;j = \{1,\dots,m\}$$ confine the relative contribution of model satellites ${\bf A}$.
Given $n$ observations of $\{x_{i},y_{i}\}$, we can construct a log-likelihood function as follows
\begin{align}\label{ahp_eq:LL}
L({\bf A}) &= \prod_{i=1}^{n}f(x_{i},y_{i}) \notag\\
		 &= \prod_{i=1}^{n}\Big\{\sum_{j=1}^{m}A_{j}\,f_{j}(x_{i},y_{i})\Big\} \notag\\
{\rm log}\, L({\bf A}) &= \sum_{i=1}^{n}{\rm log}\,\Big(\sum_{j=1}^{m}A_{j}\,f_{j}(x_{i},y_{i})\Big)
\end{align}

Maximizing ${\rm log}\;L({\bf A})$ will yield the maximum likelihood estimate ${\bf A}_{MLE}$ for ${\bf A}_{EM}$ --- our best expectation-maximization estimate for the $true$ $A_{j}$ values ${\bf A}_{T}$. This task, which can be computationally arduous, can be made tractable by adding a latent indicator, $z$, to each observed data point ($x$, $y$), to represent the model template of origin. By designating data set $\{x_{i},y_{i},{z_{i}}\}_{i=1}^{n}$ as our complete data, we can then define a complete data likelihood as
\begin{align}\label{ahp_eq:compLL}
L({\bf A}) &= \prod_{i=1}^{n}\prod_{j=1}^{m}\Big\{A_{j}\,f_{j}(x_{i},y_{i})\Big\}^{z_{ij}} \\
\ell({\bf A}) &= \sum_{i=1}^{n}\sum_{j=1}^{m}z_{ij}\,{\rm log}\,\{A_{j}\,f_{j}(x_{i},y_{i})\})
\end{align}
where $z_{ij}$ equals the {\it hard} expectation that ($x_{i},y_{i}$) comes from $j^{th}$ satellite template and $\ell({\bf A})$ is the complete data log-likelihood.

As stated above, the log-likelihood derived above can be used to obtain ${\bf A}_{EM}$ via the expectation-maximization (EM) algorithm. Starting from an initial set of guesses, ${\bf A}^{(0)}$, the algorithm iteratively steps through guesses (which are informed by the former set) until the value of the log-likelihood $\ell({\bf A})$, conditioned on the data (and within some tolerance), is maximized.  More specifically, the maximizing value of the $t^{th}$ iteration, ${\bf A}^{(t)}$, is then used as the starting value for the next run, and it continue until the likelihood changes by less than $10^{-3}$ over twenty-five iterations. Details to the implementation of this technique are shown in Appendix~\ref{ahp_app}.

We discuss how we evaluate the success of our estimates in the next section. Results from the EM estimates are discussed from \S\ref{ahp_sec:results1} onwards.

\subsection{Evaluating the success of the method}\label{ahp_sec:ATFoE}
In order to evaluate the relative success among our calculated AHPs across all halos and the success of the technique across various STS, we compare the EM estimates, ${\bf A}_{EM}$, to the known true values, ${\bf A}_{T}$. Using these values we can calculate the ``factor-of-error'' (FoE) ratio for each template EM estimate. The FoE value is defined as the maximum between ${A_{EM,j}}/{A_{T,j}}$ and ${A_{T,j}}/{A_{EM,j}}$.\footnote{This definition is chosen to obtain the most general sense of FoE statements (which are common in astronomy) such as ``the observed [generic] measurements are within a factor of 2 of theoretical predictions.'' This statement implies that observed measurements are between less-than-twice and greater-than-half of the theoretical values in question.}

One way to evaluate the fidelity of our results is to determine an average FoE ratio ($\rm{\langle FoE \rangle}$) from all FoE measured (i.e. from a given STS and halo). This $\rm{\langle FoE \rangle}$ is an average of all FoE$_{j}$, weighted by $w_{j}$, and given as
\begin{equation}\label{ahp_eq:FoE}
    \langle {\rm FoE} \rangle = \sum_{j=1}^{m}w_{j}\cdot {\rm FoE}_{j} 
\end{equation}
where $w_{j}$ represents a choice of weights for the relative importance of each template estimate and $m$ is the number of templates used. The lowest $\rm{\langle FoE \rangle}$ value indicates the best results balanced by $w_{j}$ in STS templates for each halo examined. For our primary analysis we take a mean of FoE values ($w_{j} = {m}^{-1}$) while other weights are examined in \S\ref{ahp_sec:diss}. The method of evaluation is applied to results in \S\ref{ahp_sec:results1}--\S\ref{ahp_sec:3x5}. 

\section{Results I: Accretion History Profiles in 1-D}\label{ahp_sec:results1}
In this section, we determine how accurate our satellite contribution estimates can be for our simplest STS. More explicitly, we investigate how well we can estimate the fractional contributions to a halo's construction via STS that span the stellar mass of the accreted system (i.e., its luminosity function) or its time of accretion (i.e., its stellar mass accretion history).

\subsection{Stellar mass fractions}\label{ahp_sec:LF}
As discussed in \S\ref{ahp_sec:STS}, we can construct a $true$ AHP from our model stellar halos to determine how accurately we can estimate them using the EM algorithm discussed in \S\ref{ahp_sec:EM}. Here, we examine the accuracy of our 1x5 STS estimates which are a 1-D set of 5 mass bins (as described in \S\ref{ahp_sec:STS} and shown in the top-left of Figure~\ref{ahp_fig:5x5}) --- that is to say, we evaluate how well we can recover stellar mass fraction contributions from satellites with no sensitivity to their time of accretion. 
\begin{figure}[t]
\begin{center}
   \includegraphics[width=0.475\textwidth,angle=0]{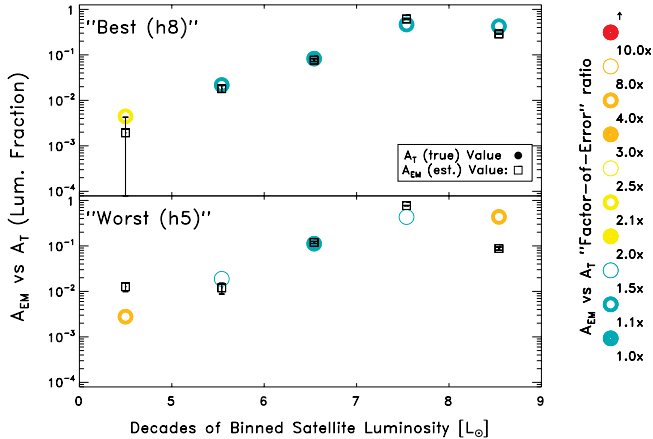}
\caption{A plot of fractional stellar mass contributions to the host halo versus the satellite's binned stellar mass for the $best$ and $worst$ EM estimates among our 11 halos (labelled h1--h11, hereafter) for 1x5 STS. Selection of these halo estimates are based on their $\rm{\langle FoE \rangle}$ values, given in respect to the number of stars (here we use $\sim10^{4}$ stars) observed. Estimates from observations (open squares) are shown for each of the five templates. Their corresponding actual values (circles) are also shown with various holes and colors that indicate the ``factor-of-error'' difference between the estimate and actual values (see legend for key). See text for discussion.}
\label{ahp_fig:haloLF}
\end{center}
\end{figure}

Figure~\ref{ahp_fig:haloLF} presents some characteristic results from our 1x5 STS analysis. The top panel legend indicates that open squares represent the ${\bf A}_{EM}$ values estimated by applying our EM analysis to observed abundances from $\sim10^{4}$ observed stars.\footnote{In a similar effort to this work, \cite{Schlaufman:2012gh} analyzed the [Fe/H] and [$\alpha$/Fe] chemical signatures of 9005 SEGUE stars in the MW (smooth) halo to ascertain the relative contributions to the accreted structure of the smooth halo finding a strong correlation between the SEGUE data and the accretion formation of MW halo analogs in $N$-body simulations at distances beyond 15 kpc from the Galactic center. Our choice of sample size demonstrates another way in which this dataset might be used.} Error bars (calculated from the Fisher information matrix) indicate the smallest possible (1$\sigma$) error values (see Appendix~\ref{ahp_app} for details). The colored circles shown represent the ${\bf A}_{T}$ (true) values while the specific colors of each circle categorize the FoE between ${\bf A}_{EM}$ and ${\bf A}_{T}$ values by the color legend to the right of the plot.  Various FoE values spanning less than 1.1 (green ``solid'' circle) to 10 or more (red ``solid'' circle) are examined. 


In the figure, two plots are chosen to display results from two representative halos (labelled by ``h'' with the designated number for the halo for short). The two halos are the $best$ (h8) and $worst$ (h5) AHP estimates as determined by their average FoE ($\rm{\langle FoE \rangle}$) values. 

Looking at our $best$ EM estimates from h8, we see that individual $A_{EM}$ produce errors that are within a factor of 2.5 or better for all template estimates using $\sim10^{4}$ observed stars. This remarkable considering that we are characterizing $\lesssim 10^{-2}$ to $10^{-3}$ of the total halo luminosity for the lowest mass bins. 

Our $worst$ EM estimates from h5 seems to reinforce the notion that this analysis provides reliable results. In this worse case scenario, most estimates are within a factor of 2 while the worse estimate (given for our most massive satellite template) is within a factor of 8. 

\subsection{Accretion time histories}\label{ahp_sec:ATF}
\begin{figure}[t]
   \centering
   \includegraphics[width=0.475\textwidth,angle=0]{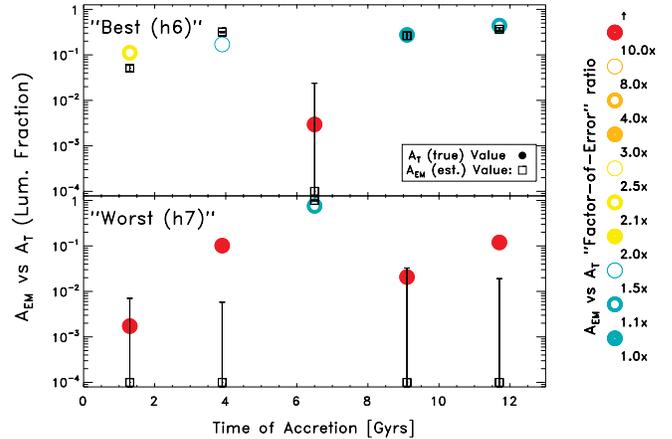}
   \caption{Figure is the same as Figure~\ref{ahp_fig:haloLF} for 5x1 STS. See text for discussion.}
   \label{ahp_fig:halo_ATF}
\end{figure}

The other principle dimension of our analysis is time. Using the same prescribed analysis above we can examine the success of estimating AHP from a 1-D set of 5 equally-spaced time bins (also described in \S\ref{ahp_sec:STS}) --- that is to say, we evaluate how well we can recover stellar mass fraction contributions from satellites with no sensitivity to their stellar masses. Figure~\ref{ahp_fig:halo_ATF} presents some characteristic results from our 5x1 STS analysis. In the figure, plots are chosen based on the same criteria used in making Figure~\ref{ahp_fig:haloLF}. 
The $best$ EM estimates from h6 reveal very different results concerning the reliability of our analysis when compared to the 1-D $mass$-$resolved$ templates results. While both the two most recent and two earliest accretion events have FoE values $\le 2.5$, the ``medieval'' accretion event has a FoE value $\gtrsim30$. Here, only the least massive accretion event has a poor FoE value.
\begin{figure*}[t]
   \centering
   \includegraphics[width=0.95\textwidth,angle=0]{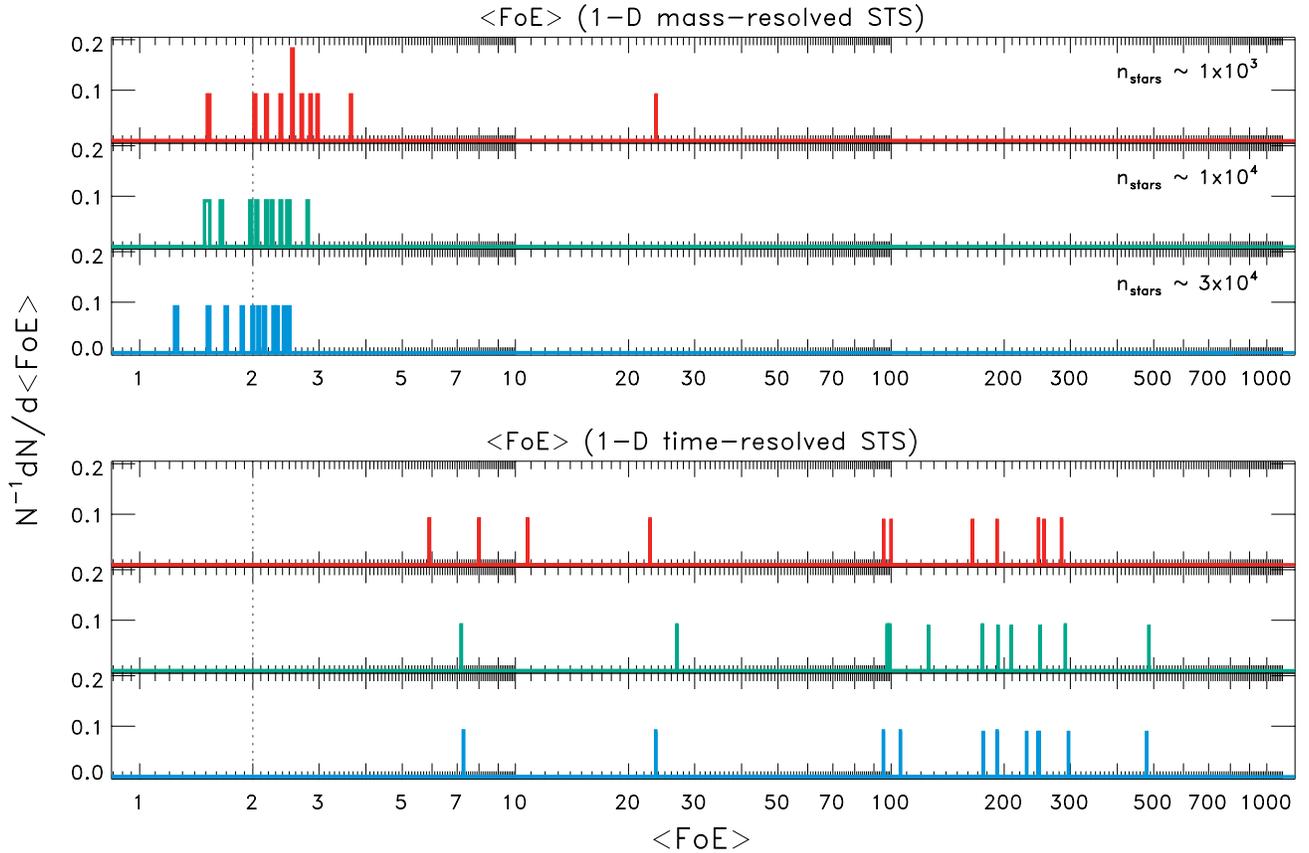}
   \caption{Figure shows panels for the frequency of $\rm{\langle FoE \rangle}$ values amongst all 11 halo for 1x5 STS (i.e. versus mass of accreted satellite; {\it top}) and 5x1 STS (i.e. versus time of accretion; {\it bottom}. Red, green, and blue histograms refer to the number of stars used to calculate the EM estimates summarized in this figure. Light-grey dotted lines indicate a $\rm{\langle FoE \rangle}$ = 2 to guide the eye when comparing the difference in results. The difference in the spread and range of $\rm{\langle FoE \rangle}$ values between the 1x5 vs 5x1 STS are striking and seem to support the notion (from Figure~\ref{ahp_fig:5x5}) that 1x5 STS retains greater distinction between its templates than the 5x1 STS do (resulting in better estimates from the 1x5 STS).}
   \label{ahp_fig:halo_FoE} 
\end{figure*}

Our $worst$ EM estimates from h7 follow a trend where all but the most massive accretion event (the medieval event in this case) have markedly poor FoE values that range from $\gtrsim20$ to $\gtrsim10^{3}$. Here, the best estimate has a FoE $\le1.5$ (i.e. with 50\% of the true value). While the estimates call into question the reliability of using multiple dimensions in $t_{acc}$ and $M_{sat}$, the overall results were already anticipated from the visual inspection of these templates in the bottom-right corner of Figure~\ref{ahp_fig:5x5}. As suggested earlier, it is likely that degeneracies in CARDs within this template set led to the poor ${\bf A}_{EM}$ estimates seen. In particular, the difference between FoE values for the medieval accretion events in h6 and h7 versus the other events comes down to the dominant accretion event templates subsuming those events that are both highly degenerate in CARD space and significantly less massive than the main event(s). As a consequence, it may appear hopeless to try to glean any information about the accretion times from 1-D estimates. This may also hold true for estimates in multiple dimensions when accretion time is treated as the dominant dimension of analysis (see \S\ref{ahp_sec:results2} for further discussion).

\subsection{Accuracy of stellar mass fractions across halo realizations: $\rm{\langle FoE \rangle}$}\label{ahp_sec:sum_FoE}
Our complete results, summarized by $\rm{\langle FoE \rangle}$, provide us with insights into the overall effectiveness of the analysis for all 11 halos. Figure~\ref{ahp_fig:halo_FoE} displays $\rm{\langle FoE \rangle}$ values for the 1x5 STS (i.e. 1-D $mass$-$resolved$; top panel) and the 5x1 STS (i.e. 1-D $time$-$resolved$; bottom panel). In both panels, each plot shows an histogram of $\rm{\langle FoE \rangle}$ values, calculated using the number of observed stars indicated in each plot, and normalized by the number of halos examined. Dotted, light-grey lines indicate a $\rm{\langle FoE \rangle}$ = 2 which indicates, by eye, the vast difference in trying to recover AHPs from 1-D mass of accreted satellites templates versus 1-D time of accretion templates. 


In our $mass$-$resolved$ (1x5 STS) estimates (top panel), we can examine the overall success of these templates and note the degree of improvement in estimates as a result of using more data points. Looking at the full panel, we can clearly see the gradual, distinct improvement in ${\bf A}_{EM}$ estimates when a larger dataset is used. The median $\rm{\langle FoE \rangle}$ (i.e., our accuracy) for each larger set of observed stars are $\sim2.55$, $\sim2.16$, and $\sim2.06$, respectively. However, its important to note that the  modest improvement between the last two datasets possibly indicates that the method is  hitting a limit due to number of templates versus the numbers of stars used.
 \begin{figure*}[t]
    \centering
    \includegraphics[width=0.95\textwidth,angle=0]{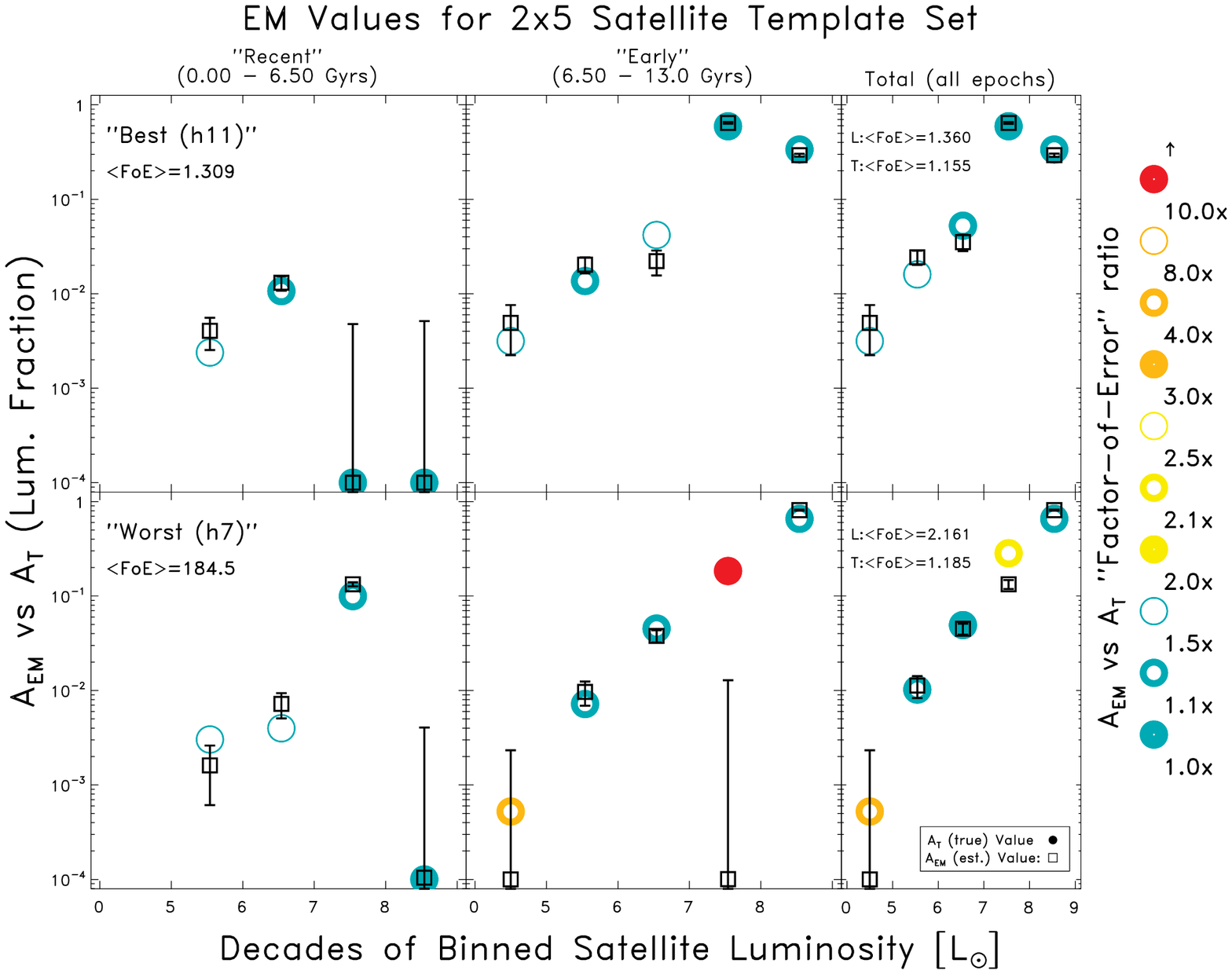} 
    \caption{Figure of 2x5 STS is similar to Figure~\ref{ahp_fig:haloLF} but first two columns shows separate sets of templates for $recent$ (0--6.5 Gyrs) and $early$ (6.5--13.0 Gyrs) accretion epochs. Final column shows totals over all time (i.e., an  ``effective'' 1x5 STS from adding corresponding estimates from both columns). Numbers labeled ``L'' and ``T'' refer to $\rm{\langle FoE \rangle}$ values calculated across satellite stellar mass and time bins, respectively.}
    \label{ahp_fig:halo_2x5}
 \end{figure*}
In our $time$-$resolved$ (5x1 STS) estimates (bottom panel), we can see that these estimates are far poorer than the estimates for the $mass$-$resolved$ estimates. In fact, the $time$-$resolved$ estimates have a median $\rm{\langle FoE \rangle}$ for each larger set of observed stars equal to $\sim100$, $\sim175$, and $\sim192$, respectively. 

Even more critical is the fact that these estimates get marginally worse with number of observations used. This suggests that there are degeneracies between templates in the set that cannot be removed with more CARD information in just two chemical abundance ratio dimensions. Conversely, these degeneracies may also suggest the inherent need for mass divisions in the STS to see differences in templates --- a possibility that motivates moving our STS to higher dimensions in the $t_{acc}-M_{sat}$ plane. In the next sections, we discuss the impact of expanding our analysis to multiple dimensions in order to achieve better estimates.

\section{Results II: Accretion History Profiles in 2-D}\label{ahp_sec:results2}
Now that we have established a baseline for estimates in our special 1-D cases, we seek to extend our search in higher dimensions of time (i.e. fixing 5 mass bins and varying our number of equally-spaced time bins). In the following subsections, we discuss our results in detail for our 2x5 and 3x5 STS (i.e. with 2 or 3 time bins), presenting insights into their success and failure.

\subsection{{\normalfont Xx5} satellite template set results}\label{ahp_sec:Xx5}
The goal of expanding our STS into higher dimensions is two-fold. First, we want to directly recover AHPs with high fidelity by dividing our $t_{acc}-M_{sat}$ plane into templates that would reveal interesting information (e.g., about the MW halo's history) when applied to real abundance observations. Second, we want to indirectly recover 1-D stellar mass functions (mass-resolved profiles) and time of accretion histories (time-resolved profiles) of our halos by summing ``like'' estimates in time or mass together (marginalization) in order to generate better accounts in 1-D than could be done directly. Our hypothesis is that allowing a finer grid in time will produce templates with less degeneracy and allow a better recovery of AHP. Of course, this must be balanced by the size of our sample and its ability to constrain the additional parameters (larger ${\bf A}_{EM}$ set) from the increased number of templates. 

\subsubsection{``Early" vs. ``recent" accretion:  {\normalfont 2x5} STS results}\label{ahp_sec:2x5}
To address our goals, we start by generating templates for our 2x5 STS which have two, evenly divided, time bins for $recent$ (0--6.5 Gyrs ago) and $early$ (6.5--13 Gyrs ago) epochs. Figure~\ref{ahp_fig:halo_2x5} displays a selection of results that reveal the success of EM estimates due to the application of our 2x5 STS. In the figure, we can once again examine the $best$ (h11), the $median$ (h2), and the $worst$ (h7) of the halo estimates using these templates. Here, in the first column of 
Figure~\ref{ahp_fig:halo_2x5}, we display the values of $\rm{\langle FoE \rangle}$ to indicate the success of estimates using $\sim10^{4}$ stars which can be compared to the our $marginalized$ results in the right-most columns. At first glance, we see that all panels indicate, by (mostly green) colors, that most estimates are within a FoE of 2. For the $best$ EM estimates (from h11), its encouraging that all FoE values are $\le2$.

However, for the $worst$ EM estimates (from h7) we see a marked decrease in the fidelity of a couple of estimates and especially for one at the high mass end. Here, we see that the $A_{T,j}$ value for the $early$ accreted $10^{7-8} M_{\odot}$ template is actually similar to its recently accreted counterpart whereas the EM estimates are very different. While the $early$ accretion event is estimated to be essentially non-existent, both the adjacent higher mass template ($early$ accreted $10^{8-9} M_{\odot}$ template) and the $recent$ accretion $10^{7-8}$ $M_{\odot}$ template have slightly higher EM estimates than there true values. The $\le50\%$ difference in FoE values is probably due to both templates subsuming the contributions from the poorly estimated $10^{8-9} M_{\odot}$ template. Given that this template is high mass and accreted early, this degeneracy is likely due to the fact that the accretion of most massive systems happens early in most of the 11 halos' histories. Since the 1515 satellites used to make the templates are comprised of 11 ensembles of accreted dwarf systems that make up the composition of our simulated halos, it is not surprising that a coarse divisions in accretion epochs lead to disparities in the fidelity of our estimates across the 6.5 Gyr divide. 

On the other hand, as indicated by our $best$ selection, it is reassuring that given the simplicity of our dwarf models, there is enough information in their CARDs to make templates that differentiate between higher mass progenitors of the halo at different epochs. This is true, despite the fact that the highest mass dwarf models show the greatest amount of degeneracy among accreted systems throughout all halos' assembly histories. Also, given the strength of current techniques to more accurately identify recent galaxy formation (e.g., color-magnitude diagrams from photometric surveys which lead to estimates for age and star formation histories and phase-space diagrams from low-res spectroscopic surveys which lead to estimates for accretion histories), it is encouraging that our technique works so well for early accretion epochs and low luminosity objects.

In the last column of Figure~\ref{ahp_fig:halo_2x5}, we present a summation of estimates across accretion epochs (shown with $\rm{\langle FoE \rangle}$ values labeled ``L'') and across binned satellite luminosities (labeled ``T'') for all epochs. Here, we confirm that a marginalization of estimates across our two epochs yields 1-D estimates with greater fidelity than its 2-D decomposition for the $worst$ EM estimates as indicated by the L-labeled $\rm{\langle FoE \rangle}$ values. More importantly, we can compare our best $worst$ values for our h7 estimates (FoE = 2.161) to the respective 1-D h8 estimates (FoE = 2.059) in Figure~\ref{ahp_fig:haloLF}. A comparison of these values shows tentative evidence that our hypothesis about gains in STS information is correct --- that the 1-D marginalizations across epochs from a 2-D STS provides on par or better estimates for 1-D AHP than does our $bona$ $fide$ 1-D STS. 
We can also compare the set of ``T''-labelled best $\rm{\langle FoE \rangle}$ values for our 1-D marginalizations across satellite luminosity bins in Figure~\ref{ahp_fig:halo_2x5} to the set of values calculated for Figure~\ref{ahp_fig:halo_ATF} (FoE = [7.168, 485.6] for our $best$ and $worst$ values, respectively). Here, we find that our estimates for our time of accretion histories improve substantially overall, and dramatically when comparing our $best$ and $worst$ AHP estimates. The next two sections address whether these improvements are ubiquitous as we increase the resolution of our STS in the accretion time dimension.

\begin{figure*}[thp] 
    \centering
    \includegraphics[width=0.95\textwidth,angle=0]{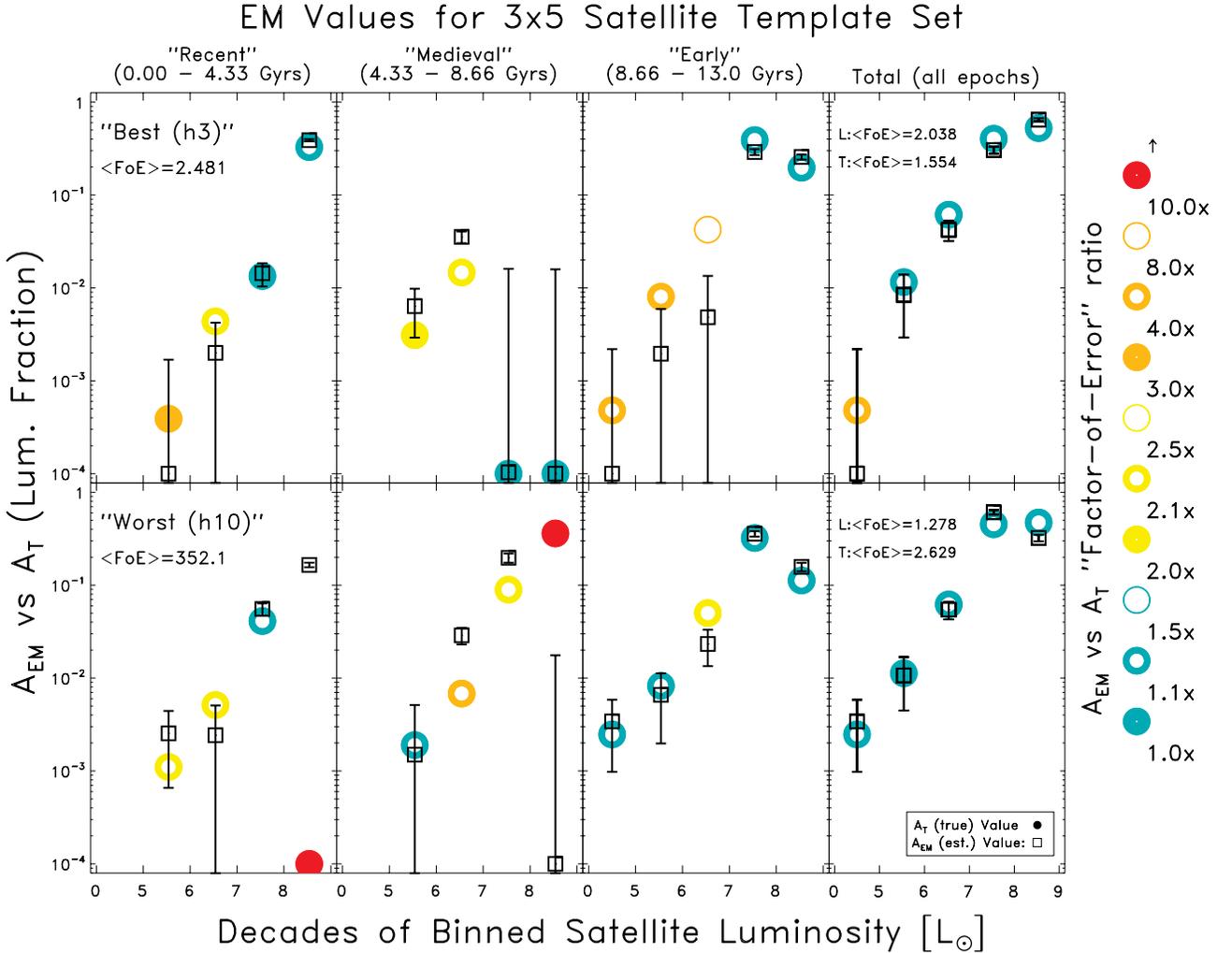} 
    \caption{Figure of 3x5 STS is similar to Figure~\ref{ahp_fig:halo_2x5} but includes an addition column for an intermediate $medieval$ accretion epoch.}
    \label{ahp_fig:halo_3x5}
 \end{figure*}  
 
\subsubsection{``Medieval" accretion:  {\normalfont 3x5} STS results}\label{ahp_sec:3x5} In order to further test our ability to estimate AHPs, we seek to increase our accretion time resolution (by adding an intermediate ``medieval" accretion epoch), with the hopes that greater information from an expanded STS will lead to better AHP estimates. 

 \begin{figure*}[thp] 
    \centering
    \includegraphics[width=0.98\textwidth,angle=0]{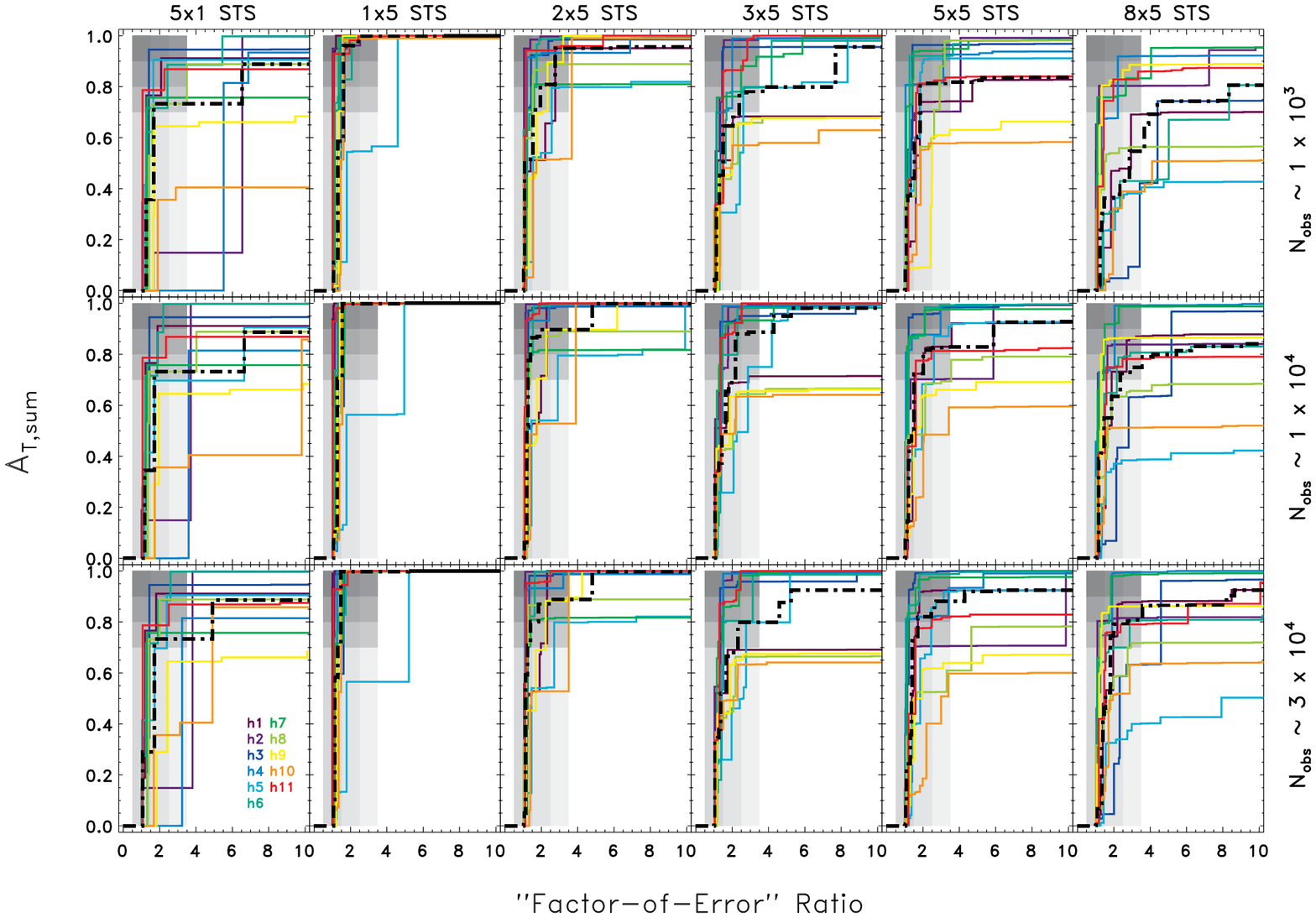} 
    \caption{Figure displays six STS-derived plots of $A_{T,sum}$($\le$ FoE) for all 11 halos demonstrating another benchmark for our CARD analysis for deriving the AHPs of our halos. Columns represent results for listed STS estimates. Rows represent estimates derived from a certain number of observed stars which are labeled at the right edge of each row. Shaded areas in each plot guide the eye to FoE estimates of $\sim2-3$ or better which primarily indicate estimates that cover $A_{T,sum}\gtrsim 70\%$. Individual solid colored lines represent each of the 11 haloes used in the study. Colored labels for the halos are shown in the bottom-left of the figure. Black dot-dashed CDF represent the median of all 11 halos vs. FoE values.} 
    \label{ahp_fig:cdf_FoE}
 \end{figure*} 
 
Figure~\ref{ahp_fig:halo_3x5} shows our $best$ and $worst$ 3x5 STS results. The $\rm{\langle FoE \rangle}$ values between the $best$ and $worst$ EM estimates show a substantial decrease in quality. It is immediately apparent (from color) that individual estimations fared significantly worse than they did in the 2x5 STS selections of Figure~\ref{ahp_fig:halo_2x5}. Also, by inspection, the $medieval$ epoch yields the worst estimates overall. Similar to Figure~\ref{ahp_fig:halo_2x5}, early epoch estimates of Figure~\ref{ahp_fig:halo_3x5} are the most accurate. The overall decrease in performance from our 2x5 to 3x5 STS is likely due to the degeneracy in CARD space between some adjacent templates in the 3x5 STS (e.g., see Figure~\ref{ahp_fig:5x5} for illustration of this effect) and across accretion time for the higher luminosity templates. For example, if we look across the $recent$ and $medieval$ epochs for our $worst$ EM estimate selection, we can see that there are degeneracies in the estimates for the highest stellar mass bins ($10^{8-9} M_{\odot}$). These degeneracies are due to the increasing similarities between chemical model tracks of more massive (and luminous) dwarf satellite models. Such degeneracies can lead to the satisfaction of estimates across all epochs by one individual template (e.g., h7 from Fig.~\ref{ahp_fig:halo_ATF}), by distributing the luminosity fraction amongst co-degenerate templates (e.g., h7 from Fig.~\ref{ahp_fig:halo_2x5}), or by swapping estimates across adjacent epochs (e.g., h10 from Fig.~\ref{ahp_fig:halo_3x5}). However, it appears that a clear separation in accretion epochs for the same stellar mass bins possibly reduces degeneracies between them (as seen for the $best$ (h3) estimates). 


If we look at the final column for our 1-D marginalizations from the 2-D 3x5 STS, we once again see improvements in $\rm{\langle FoE \rangle}$ values in comparison to Figures~\ref{ahp_fig:haloLF} and~\ref{ahp_fig:halo_ATF} (e.g., look at ``L'' and ``T'' values for all selections versus uniformly-weighed values in Fig.~\ref{ahp_fig:FoEwghts} of \S\ref{ahp_sec:diss}). While improvements were anticipated, it is still surprising, given the relative lack of success for individual 3x5 STS templates, that marginalization of the $worst$ 3x5 STS leads to 1-D estimates that offer an improvement over the 2x5 STS marginalized 1-D estimates.  In this case, some inaccuracies due to degeneracies across epochs are mitigated by summation over accretion epochs. Consequently, improvements to our marginalized $mass$-$resolved$ 1-D estimates arise from an increase in the STS epoch resolution. Presumably, the better estimates would originate directly from improved individual epoch estimates. However, poor individual estimates due to degeneracies within the same stellar mass bins refute this idea. Indeed, it is more likely that improvements to our epoch resolution led to better estimates indirectly, by not decreasing the degeneracies between adjacent epochs, but rather decreasing degeneracies between adjacent stellar mass bins. While the effects described above are certainly taking place, it is still unclear from Figures~\ref{ahp_fig:haloLF}, ~\ref{ahp_fig:halo_ATF}, ~\ref{ahp_fig:halo_2x5}, and~\ref{ahp_fig:halo_3x5} whether these improvements remain across all 11 halos. In the next section we examine the $\rm{\langle FoE \rangle}$ values as ensembles across the 11 halos to determine the overall success of recovering AHPs given our STS.
 \begin{figure*}[thp] 
    \centering
    \includegraphics[width=0.95\textwidth,angle=0]{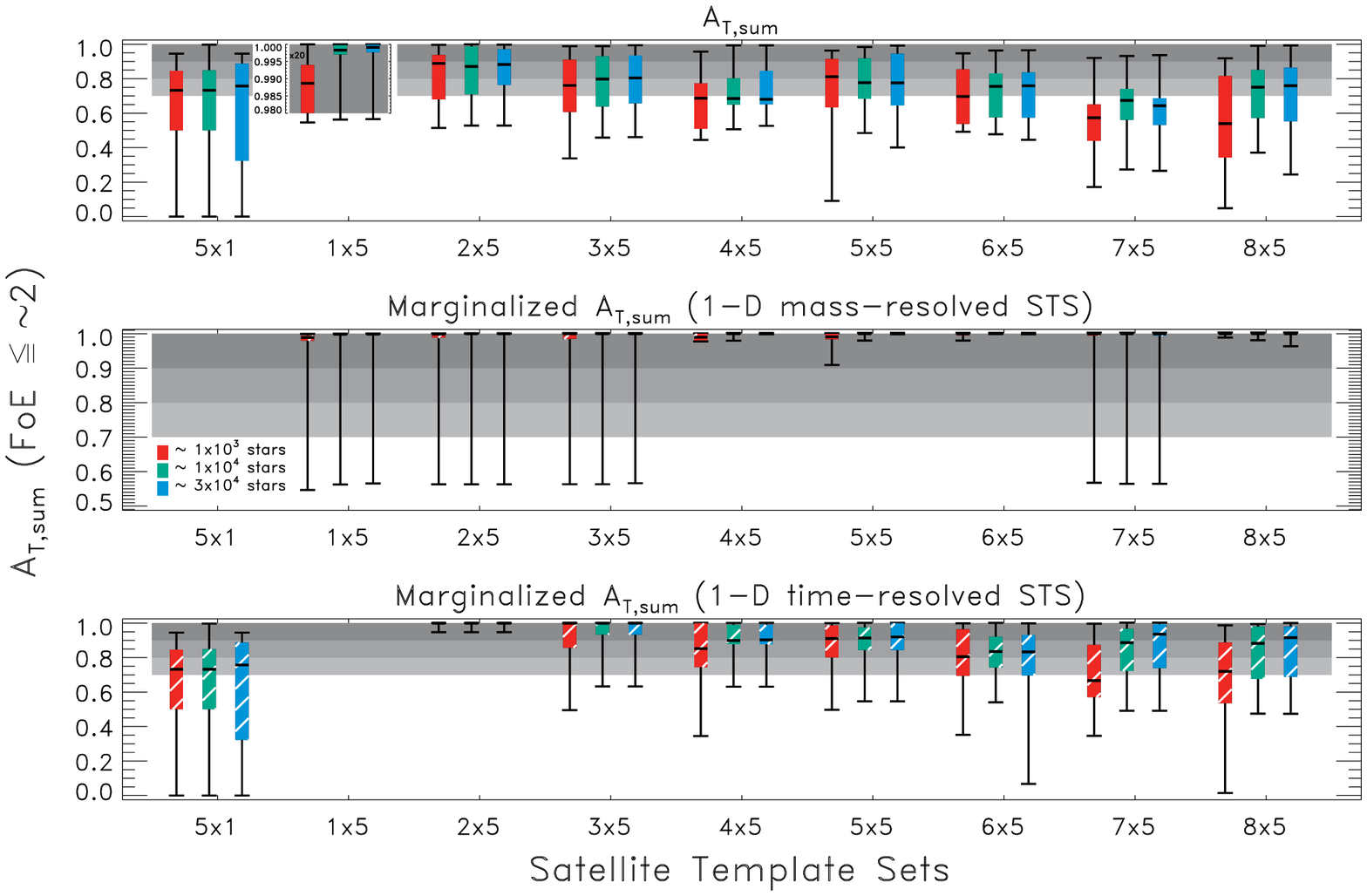} 
    \caption{Figure shows box-and-whisker plots of $A_{T,sum}$(FoE $\lesssim2$) for full STS (top), marginalized 1-D $mass$-$resolved$ STS (middle), and marginalized $time$-$resolved$ (bottom) using all STS examined for our 11 halos.  The median values of $A_{T,sum}$ for all 11 halos are shown as a black line across every box. The $25^{th}$ and $75^{th}$ percentiles of the distribution are shown as the lower and upper bounds of the each box, respectively. Whiskers designate the minimum and maximum values for $A_{T,sum}$ values in the distributions shown. Each box has a color that refers to the number of stars identical to the colors used in Figure~\ref{ahp_fig:halo_FoE}.  $Top$: Boxes in the top panel (solid colors) refer to the genuine $A_{T,sum}$ values for each respective STS. $Middle$ and $Bottom$: ``Marginalized'' boxes (striped colors) refer to the $A_{T,sum}$ values calculated from the sum across the mass (time) dimension of templates into an effective 1x5 (5x1) template (e.g., see Figures~\ref{ahp_fig:halo_2x5} and~\ref{ahp_fig:halo_3x5}). 1x5 STS ($mass$-$resolved$) $A_{T,sum}$ values derived from marginalizing over time-binned estimates are shown in the middle panel while 5x1 STS ($time$-$resolved$) $A_{T,sum}$ values derived from marginalizing over mass-binned estimates are shown in the bottom panel. Increasingly darker grey bands spanning all STS  (for $70\% \le A_{T,sum} \le 100\%$) are shown to highlight the success of our estimates.}
    \label{ahp_fig:comp_FoE}
 \end{figure*} 
 
\subsection{Comparison of results across all STS}\label{ahp_sec:all}
In this section we compare results from all our simulated halos and the templates we constructed. Using FoE values (see \S\ref{ahp_sec:ATFoE}) we can determine a cumulative distribution function (CDF) of FoE values with respect to ${\bf A}_{T}$ for each STS used. The CDF values described above (which we call $A_{T,sum}$) indicate the fraction of the total stellar halo mass we can identify within a given FoE value. 

First, we construct $A_{T,sum}$ values in Figure~\ref{ahp_fig:cdf_FoE} for six of our 10 STS. Each plot frames the recovery of AHPs in terms of the level of accuracy (i.e., FoE) at which we can characterize a certain portion ($A_{T,sum}$) of the total luminous stellar content of the halos examined. Once again differences in the fidelity of our estimates between 5x1 and 1x5 STS are clearly shown with a median $A_{T,sum}$ (fraction recovered) with a FoE $\lesssim2$ being $\simeq 73\%$ and 95--99\%, respectively. Characterizing the success of the method overall, we find that the median $A_{T,sum}$(with FoE $\lesssim2$) across most STS is $\simeq 70\%$ or better. It is evident from the STS shown in Figure~\ref{ahp_fig:cdf_FoE} that EM estimates fair poorly when applied to certain halo realizations. We discuss possible causes for the often poorer estimates of a few halos in \S\ref{ahp_sec:diss}.

Figure~\ref{ahp_fig:comp_FoE} displays another way we can summarize our results with the utilization of $A_{T,sum}$ and FoE. In the three panels, box-and-whisker plots illustrate the median and shape of the distribution of $A_{T,sum}$ values calculated for estimates with FoE $\lesssim2$ amongst all 11 halos.\footnote{The actual chosen cutoff here for FoE values is $\le2.25$. Given that this research is presented as a proof-of-concept, we wanted to capture FoE values that were consistent with a FoE = 2. Since such a cutoff is arbitrary, the reader is free to reexamine the selected columns of Figure~\ref{ahp_fig:cdf_FoE} and reconstruct $A_{T,sum}$ estimates for different FoE cutoff values.} 
The top panel displays similar information to the results shown in Figure~\ref{ahp_fig:cdf_FoE}. The middle and bottom panels show both genuine and marginalized estimates for the 1x5 STS accreted mass functions and the 5x1 STS accretion time histories, respectively. 

In the top panel, $A_{T,sum}$(FoE $\lesssim2$) is plotted, as a color box, for all STS examined. Here, as in Figure~\ref{ahp_fig:halo_FoE}, the color refers to the respective number of observations used (as indicated in the plot legend). In the plot, we see that our best median $A_{T,sum}$ values are given by the 1x5 and 2x5 STS while the worse values are given by 5x1 and 7x5 STS. The average among the best and worst $A_{T,sum}$ values across all STS examined and for increasing number of stellar observations are $\sim0.96-0.98$ and $\sim0.29-0.41$, respectively. The average median $A_{T,sum}$ values across all STS examined and for increasing number of stellar observations are 0.742, 0.783, and 0.785. This means that on average our FoE are $\lesssim2$ for at least $\sim75$\% of the total halo stellar mass (i.e. $A_{T,max}=0.75$) observed. 

Marginalized values, which are defined in \S\ref{ahp_sec:2x5}, are useful for evaluating any gains that may potentially arise due to better time (or mass) resolution.  More precisely, any information about templates that is lost or gained should generally result in a corresponding rise or drop in $\rm{\langle FoE \rangle}$ and thus appear as an increase in $A_{T,sum}$(FoE $\lesssim2$). As a reference, a grey bar is placed in each panel to indicate a region where the $A_{T,sum}$(FoE $\lesssim2$) values range from 70\% to 100\% (from bottom to top).
 
The middle panel shows our $mass$-$resolved$ marginalized values (summed over accretion time bins) for 8 of the 9 STS (with 5x1 omitted because its value is not applicable in this context). The plot shows an across-the-board increase in $A_{T,sum}$(FoE $\lesssim2$) values (i.e., a general drop in all STS $\rm{\langle FoE \rangle}$ values) measured for a recovery of the total stellar mass function. The improvement in $\rm{\langle FoE \rangle}$ despite the tendency for individual FoE STS values to increase with an increase in the number of templates used indicates that significant gains were made by using a larger template set for the specific purpose of generating more accurate estimates of a halo's total stellar mass function (via marginalization).

The bottom panel shows our $time$-$resolved$ marginalized values (summed over mass bins) for 8 of the 9 STS (with 1x5 also omitted because its value is not applicable in this context). In this case, the plot shows a descending trend in $A_{T,sum}$(FoE $\lesssim2$) values with larger STS (i.e., a generally ascending rise in $\rm{\langle FoE \rangle}$ values with increasing STS size) measured for a recovery of the total accretion time history. Despite the decrease $A_{T,sum}$(FoE $\lesssim2$) values, these values remain relatively good (above 70\% for $A_{T,sum}$ values above the bottom 50\% margin) up to our 6x5 STS. Indeed, all $time$-$resolved$ marginalized values show a significant improvement in accretion time histories over the history given by the 5x1 STS.  Overall, the results show that we could expect to recover accretion time histories using the EM algorithm given that we use reasonable templates.

Results shown in Figure~\ref{ahp_fig:cdf_FoE} and Figure~\ref{ahp_fig:comp_FoE} prove that even with the simplest template divisions, we could, with the appropriate dataset, recover the accretion history of the MW halo. To that point, we find that these STS EM estimates can recover the total contributions from accreted systems (templates) of similar mass (i.e. halo luminosity function) to within a factor of 1.02 ($\le$2\% of the $true$ value) for most of the 11 halos. Separately, the EM algorithm can determine the mass fractions within accretion times to within a factor of $\gtrsim4$ for at least 90\% of the halo's total stellar mass. Both results present encouraging prospects for recovering the accretion history of the MW halo from current and near-future data collections.

\section{Discussion}\label{ahp_sec:diss}
In the following discussion, we examine the statistical reliability (or robustness) of the EM algorithm when applied to our models and simulated data. We also explore what masses the current approach is most sensitive to and discuss implications for future work.

\subsection{Reliability}
\begin{figure}[t] 
    \centering
    \includegraphics[width=0.475\textwidth,angle=0]{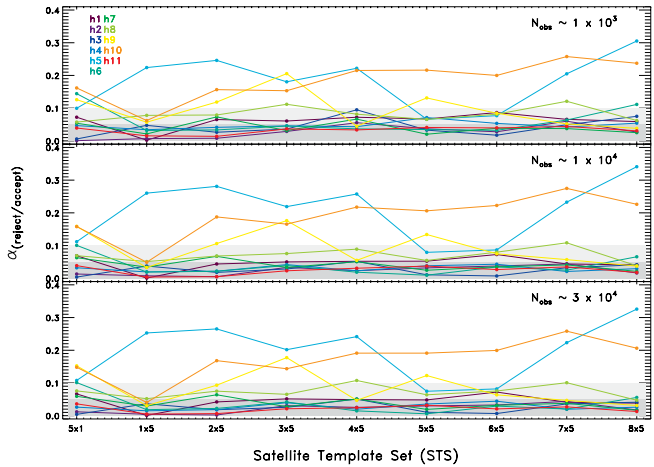} 
    \caption{Figure shows the $\alpha$-level threshold for accepting or rejecting the null hypothesis that suitable AHP templates were used in estimating ${\bf A}_{T}$ values. Colors represent results for the 11 halos examined and panels compare results for the approximate number of stars observed. See text for discussion.}
    \label{ahp_fig:lltest}
 \end{figure}

We can test the statistical robustness of the EM algorithm's application to our simulated halos by performing a {\it likelihood ratio test} on the results of our analysis. By determining the true (${\bf A}_{T}$) and respective ${\bf A}_{EM}$ likelihood values from each application of STS to our halos via the EM algorithm, we can calculate a $\chi^{2}$-statistic defined by the following equation 
\begin{equation}\label{ahp_eq:lltest}
\chi^{2} = -2\; ln\Big(\frac{\lambda_{T}}{\lambda_{EM}}\Big) 
\end{equation}
where $\lambda_{T}$ and $\lambda_{EM}$ are the likelihoods for ${\bf A}_{T}$ and ${\bf A}_{EM}$ values, respectively. One can then reject the assumption that the true AHP templates are well-approximated by the STS used if the $\chi^{2}$-value from Eqn.~\ref{ahp_eq:lltest} is larger than the $\chi^{2}$-percentile values given $k$ degrees-of-freedom ($k$ =  $m_{EM}$ - $m_{T}$)\footnote{Hence $k$ equals the number of templates in a STS estimate ($m_{EM}$) minus the number of those templates that are actually occupied in the true AHP ($m_{T}$).} and a confidence level denoted by $\alpha$. Figure~\ref{ahp_fig:lltest} shows the maximum $\alpha$-value one can assume for a $\chi^{2}$-distribution before you have to reject the assumption that suitable AHP templates are chosen. For example, an $\alpha = 0.05$ corresponds to a confidence that 95\% of all samples taken of a given size are well characterized by the STS in use. Here, we find that all sample sizes and STS used, halos 5, 9, and 10 are by far the worst characterized halos by our STS divisions. For most STS used, these halos are ill-matched to the generic STS created in our division scheme and therefore challenge the robustness of this method. Such challenges need to be address before this method can be utilized to model the AHP of the MW halo. The solution resides in the development and incorporation of sufficiently realistic models of dwarf CARDs into this method --- a goal that will be addressed in future work.
 
\subsection{Sensitivity to different mass bins}
\begin{figure}[t] 
    \centering
    \includegraphics[width=0.475\textwidth,angle=0]{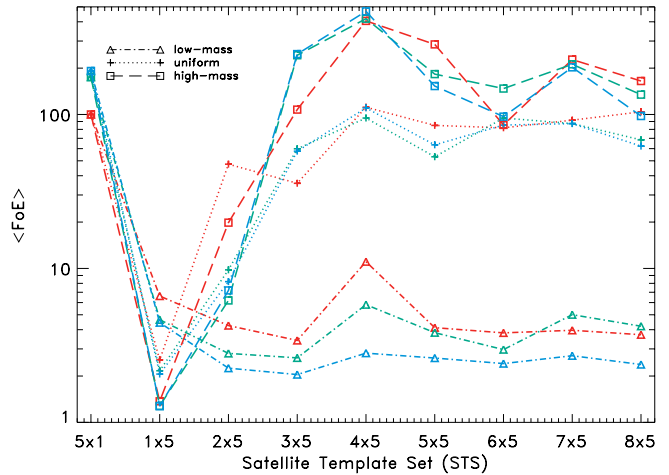} 
    \caption{Figure shows $\rm{\langle FoE \rangle}$ values for different template weights. The various colors refer to the approximate number of stars used as indicated in Fig.~\ref{ahp_fig:comp_FoE}. Weights are listed in the figure legend. See text for discussion.}
    \label{ahp_fig:FoEwghts}
 \end{figure}

Another consideration in assessing the reliability of our method is to determine how well it uncovers AHPs based on the satellite mass regime we are interested in. Taking Eqn.~\ref{ahp_eq:FoE} from \S\ref{ahp_sec:ATFoE}, we can calculate $\rm{\langle FoE \rangle}$ values with different weights --- i.e., uniform (mean), low-mass preferred, or high-mass preferred --- based on what satellite population(s) one prefers to recover. Figure~\ref{ahp_fig:FoEwghts} shows the median $\rm{\langle FoE \rangle}$ amongst all halos for each STS used. The same colors from Figure~\ref{ahp_fig:comp_FoE} are used indicate the number of stars used for the analysis and symbols and corresponding lines refer to the type of weighting used (see figure legend). Uniformly-weighted $\rm{\langle FoE \rangle}$ values are weighted by $m^{-1}$ (i.e. by the number of templates used) and identical to the weighting used for the main results of this paper. Weights that emphasize more accuracy in low- or high-mass satellite AHPs are weighted by the corresponding upper bin mass limits and their reciprocals, respectively. 

In the figure, we can see that $\rm{\langle FoE \rangle}$ values for low-mass satellite recovery fair the best whereas uniform and high-mass satellite recovery-emphasized weights are a factor of $\gtrsim10$ in all but the three smallest template sets. In other words, when one emphasizes the accurate recovery of low-mass satellites, the weighting favors templates with lower FoE values which yields lower overall $\rm{\langle FoE \rangle}$ values. This result further clarifies the immediate strengths of the method: its adept at differentiating between accreted dwarfs of low-mass in CARD-space due to the lack of degeneracies in their occupied region of space. Meanwhile, its clear that while degeneracies exist in the CARD-space occupied by high-mass satellites and larger STS, we are encouraged by the fact that the introduction of more templates can significantly decrease degeneracies in only two dimensions of CARD-space.

\subsection{Future Prospects}
It is clear from both our results and our reliability tests that the current method fails often for three of the 11 halo simulations. From our examination of these three problematic halos we find that all of them show predominately early accretion of massive dwarf galaxies with integrated CARDs that appear to be highly-degenerate when compared to the other eight halos AHP CARDs examined. 
To address the degeneracies that exist (particularly among high-mass systems) we posit that differences between mass-dependent (nucleosynthetic) yields for different nucleosynthetic sites and elements groups \citep[e.g., see][]{lee13} can be exploited to greatly reduce or remove such degeneracies by expanding the CARD-space basis set. 

For example, we only looked at two dimensions in CARD space whereas more recent work on ``chemical tagging'' expands the number of dimensions available by establishing the best chemical abundance signatures to pursue in chemical abundance space in order to optimize survey efforts (e.g., the GALAH survey). One way to optimize our surveys for searches in chemical abundance space is to prioritize spectroscopic observations for elements that confer the greatest amount of distinction between systems with different origins. To this end principle component analysis (PCA) was used by \cite{ting12} to identity and rank the $6-9$ most distinguishing elements in chemical abundance space. In their work, the chemical abundance space of various parts of both the galaxy and the galactic neighborhood were examined to determine the best elements to observe in order to decipher their galactic chemical evolution. A CARD-space basis set derived from various combinations of these elements are likely to offer the breaks in degeneracies that we require.

%
%
%
%

\section{Summary}\label{ahp_sec:conc}
In our investigation to determine the efficacy of recovering the accretion history of the MW halo, we used simulated halo data from the \cite{bullock05} MW halo simulations. Our approach required the CARDs of [$\alpha$/Fe] and [Fe/H] for the 11 simulated realizations for accretion-grown halos, observed samples of stars from those simulations, and CARD templates of accreted dwarfs models in the simulations. From this assortment of data we were able to apply a statistical algorithm (the EM algorithm) which utilizes the model templates with those $observed$ stars to disentangle the accretion history of our simulated halos. 

To evaluate the success of our estimates, we examined relationships between a measure of accuracy, the FoE, and a measure of the maximum fraction of the halo's stellar mass that is characterized by this level of accuracy which we call $A_{T,sum}$. 

In our analysis, we employed (equally-partitioned) STS as model sets for our generative mixtures (i.e., the simulated halos). The first test of our templates involved 1-D STS which were composed entirely of either stellar mass or accretion time partitions. In the case of our 1-D $mass$-$resolved$ STS, the EM algorithm estimates for individual templates were made to within a factor of 8 (in the worst case) for halo 5 and were within a factor $1.5-2.5$ or better for most mass bins. However, in the case of our 1-D $time$-$resolved$ STS, results were considerably less accurate, with approximately half of the individual templates being off by a factor of 10 or more. In this case, it is important to note that the bulk of these poor estimates occurred for bins containing the least amount of accreted mass. This outcome was not unexpected, but it stands in sharp contrast to estimates that resulted from our $mass$-$resolved$ case. In both cases, we also examined the effect of increasing our datasets from one thousand to thirty thousand stellar chemical abundance observations. While we found that an increase in our data generally led to better estimates from our $mass$-$resolved$ templates no improvement was seen for estimates from our $time$-$resolved$ templates. These results lead us to examine what, if any, improvements could be made in our EM estimates by expanding our STS into two dimensions of accretion time and mass and increasing the number of templates used.

In examining the use of the 2-D STS in EM algorithm estimations, we find that these template sets provided more accurate estimates in general. More precisely, we find that our 2x5 STS could be used to furnish remarkably good AHP estimates --- meaning that we could easily recover a tally of satellites that fell in $recently$ versus those that fell in more than 6.5 Gyrs ago. It is clear that in this dichotomous evaluation mode, the EM algorithm can easily detect distinction between previous satellites that were accreted from 6.5 Gyrs ago to now and those satellites that accreted prior to that time using only two dimensions in chemical abundance space. Also, we find that in the case where we try to estimate an $early$, $medieval$, and $recent$ accretion history --- our 3x5 STS tests --- the EM estimates do fairly well too. In some cases it was apparent from our 2-D STS figures (for our 3x5 STS in particular) that degeneracies between templates in a set were possibly degrading our EM estimates and perhaps limiting the potential for this technique. However, despite such degeneracies, we find that we can improve our 1-D recovery of both the mass accretion history (functionally similar to mass/luminosity functions) and the accretion time history (a coarse account of mass growth of the halo over time) by marginalizing estimates across templates in the appropriately related dimension. Thus, we are confident that at the very least this technique can be used, albeit carefully, to produce fairly accurate estimates for 1-D accretion mass or mass growth functions for the MW halo.

Finally, we compare our tests for all 2-D STS. We find three interesting features that reflect the technique's potential. These features are: (1) fairly accurate estimates for AHPs across most STS used (2) consistent or improved 1-D $mass$-$resolved$ $A_{T,sum}$ values from 1-D marginalization over an increase in the number of templates used, and (3) a substantial overall improvement in the marginalized $time$-$resolved$ $A_{T,sum}$ values across all STS used over the 1-D 5x1 STS values.  From these features we conclude that, on average, we can recover the bulk of accreted dwarfs' relative contributions to the halo's accretion history by mass, to within a factor of $\sim2$. Despite this fact, many individual templates (especially our lower mass bin templates) can produce estimates that are far less accurate than estimates given for the main stellar mass contributors to the halo. This is likely due to degeneracies among templates belonging to same STS and relative contributions of these objects to the general star count of halo. These issues that can be addressed by carefully selecting which observed stars are to be included in the data sample and by expanding the chemical abundance space basis set to better disentangle the individual star formation histories of the previously accreted dwarf satellites in our halos (or our Halo).

Lastly, in spite of the demonstrated drawbacks involving degeneracies between individual templates, we find that, remarkably, it is possible to improve 1-D mass function predictions (as a function of accreted satellite mass or accretion time) simply by increasing the number of partitioned time bins (templates) used for EM estimates and then marginalizing over those estimates in either stated dimension. This result means that at the very least it is possible to extract, e.g., accurate luminosity functions with estimates that clearly improve with better resolution in our $t_{acc}-M_{sat}$ plane. Further investigation of this result will be pursued in the near future.

\section{Conclusions}
In conclusion we note the following implications of our study:
\begin{itemize}
\item Our proof-of-concept is verified --- recovering halo accretion histories using their CARD information works (and works well for a certain level of detail)
\item In particular, even when applying our method to only 2-D CARD-space we appear to be sensitive to:
\item[] - early accretion events (regions where information in phase-space has phase-mixed away)
\item[] - low luminosity dwarfs (objects we cannot see $in$-$situ$ because they are too faint)
\item There $are$ degeneracies in 2-D CARD-space, particularly amongst high mass accreted dwarfs
\item However, since we only looked in 2-D and there are prospects of 10's of thousands of stars with $> 6$ independent chemical dimensions it is very important to pursue this method of approach further
\end{itemize}
Finally, given these implications we are compelled to generate more realistic templates  from chemical evolution models in higher dimensions and test them against existing dwarf data. It is the hope that by validating the fidelity of such templates, we could, in turn, employ these templates in our method to produce a detailed account of the accretion history of the MW halo.\\ \\

DML would like to give thanks to his dissertation thesis committee for their helpful comments and support in the writing of this paper. KVJ and DML would also like to give thanks to James Bullock, Brant Robertson and Andreea Font for the collaboration that developed the numerical data sets used in this work. This work was supported by the ``973 Program'' 2014 CB845702; the Strategic Priority Research Program ``The Emergence of Cosmological Structures'' of the Chinese Academy of Sciences (CAS; grant XDB09010100). DML was also supported by NSF grants AST-0806558 and AST-1107373.\\

\bibliography{reading1,allrefs}

\begin{thebibliography}{45}
\expandafter\ifx\csname natexlab\endcsname\relax\def\natexlab#1{#1}\fi

\bibitem[{{Belokurov} {et~al.}(2006){Belokurov}, {Zucker}, {Evans}, {Gilmore},
  {Vidrih}, {Bramich}, {Newberg}, {Wyse}, {Irwin}, {Fellhauer}, {Hewett},
  {Walton}, {Wilkinson}, {Cole}, {Yanny}, {Rockosi}, {Beers}, {Bell},
  {Brinkmann}, {Ivezi{\'c}}, \& {Lupton}}]{belokurov07a}
{Belokurov}, V. {et~al.} 2006, \apjl, 642, L137

\bibitem[{{Bland-Hawthorn} \& {Freeman}(2004)}]{bland-hawthorn04}
{Bland-Hawthorn}, J., \& {Freeman}, K.~C. 2004, Publications of the
  Astronomical Society of Australia, 21, 110

\bibitem[{{Bland-Hawthorn} {et~al.}(2010){Bland-Hawthorn}, {Karlsson},
  {Sharma}, {Krumholz}, \& {Silk}}]{bland-hawthorn10}
{Bland-Hawthorn}, J., {Karlsson}, T., {Sharma}, S., {Krumholz}, M., \& {Silk},
  J. 2010, \apj, 721, 582

\bibitem[{{Bonifacio} {et~al.}(2004){Bonifacio}, {Sbordone}, {Marconi},
  {Pasquini}, \& {Hill}}]{bonifacio04}
{Bonifacio}, P., {Sbordone}, L., {Marconi}, G., {Pasquini}, L., \& {Hill}, V.
  2004, \aap, 414, 503

\bibitem[{{Bullock} \& {Johnston}(2005)}]{bullock05}
{Bullock}, J.~S., \& {Johnston}, K.~V. 2005, \apj, 635, 931

\bibitem[{{Cayrel} {et~al.}(2004){Cayrel}, {Depagne}, {Spite}, {Hill}, {Spite},
  {Fran{\c c}ois}, {Plez}, {Beers}, {Primas}, {Andersen}, {Barbuy},
  {Bonifacio}, {Molaro}, \& {Nordstr{\"o}m}}]{cayrel04}
{Cayrel}, R. {et~al.} 2004, \aap, 416, 1117

\bibitem[{{De Silva} {et~al.}(2007){De Silva}, {Freeman}, {Asplund},
  {Bland-Hawthorn}, {Bessell}, \& {Collet}}]{de-silva07}
{De Silva}, G.~M., {Freeman}, K.~C., {Asplund}, M., {Bland-Hawthorn}, J.,
  {Bessell}, M.~S., \& {Collet}, R. 2007, \aj, 133, 1161

\bibitem[{{Efstathiou} {et~al.}(1985){Efstathiou}, {Davis}, {White}, \&
  {Frenk}}]{efstathiou85}
{Efstathiou}, G., {Davis}, M., {White}, S.~D.~M., \& {Frenk}, C.~S. 1985,
  \apjs, 57, 241

\bibitem[{{Eggen} {et~al.}(1962){Eggen}, {Lynden-Bell}, \& {Sandage}}]{eggen62}
{Eggen}, O.~J., {Lynden-Bell}, D., \& {Sandage}, A.~R. 1962, \apj, 136, 748

\bibitem[{{Font} {et~al.}(2006){Font}, {Johnston}, {Bullock}, \&
  {Robertson}}]{font06}
{Font}, A.~S., {Johnston}, K.~V., {Bullock}, J.~S., \& {Robertson}, B.~E. 2006,
  \apj, 638, 585

\bibitem[{{Freeman} \& {Bland-Hawthorn}(2002)}]{freeman02}
{Freeman}, K., \& {Bland-Hawthorn}, J. 2002, \araa, 40, 487

\bibitem[{{Fulbright}(2002)}]{fulbright02}
{Fulbright}, J.~P. 2002, \aj, 123, 404

\bibitem[{{Geisler} {et~al.}(2005){Geisler}, {Smith}, {Wallerstein},
  {Gonzalez}, \& {Charbonnel}}]{geisler05}
{Geisler}, D., {Smith}, V.~V., {Wallerstein}, G., {Gonzalez}, G., \&
  {Charbonnel}, C. 2005, \aj, 129, 1428

\bibitem[{{Geisler} {et~al.}(2007){Geisler}, {Wallerstein}, {Smith}, \&
  {Casetti-Dinescu}}]{geisler07}
{Geisler}, D., {Wallerstein}, G., {Smith}, V.~V., \& {Casetti-Dinescu}, D.~I.
  2007, \pasp, 119, 939

\bibitem[{{Helmi} \& {de Zeeuw}(2000)}]{helmi00}
{Helmi}, A., \& {de Zeeuw}, P.~T. 2000, \mnras, 319, 657

\bibitem[{{Ibata} {et~al.}(1994){Ibata}, {Gilmore}, \& {Irwin}}]{ibata94}
{Ibata}, R.~A., {Gilmore}, G., \& {Irwin}, M.~J. 1994, \nat, 370, 194

\bibitem[{{Ivans} {et~al.}(1999){Ivans}, {Sneden}, {Kraft}, {Suntzeff},
  {Smith}, {Langer}, \& {Fulbright}}]{ivans99}
{Ivans}, I.~I., {Sneden}, C., {Kraft}, R.~P., {Suntzeff}, N.~B., {Smith},
  V.~V., {Langer}, G.~E., \& {Fulbright}, J.~P. 1999, \aj, 118, 1273

\bibitem[{{Johnson} {et~al.}(2006){Johnson}, {Ivans}, \& {Stetson}}]{johnson06}
{Johnson}, J.~A., {Ivans}, I.~I., \& {Stetson}, P.~B. 2006, \apj, 640, 801

\bibitem[{{Jonsell} {et~al.}(2005){Jonsell}, {Edvardsson}, {Gustafsson},
  {Magain}, {Nissen}, \& {Asplund}}]{jonsell05}
{Jonsell}, K., {Edvardsson}, B., {Gustafsson}, B., {Magain}, P., {Nissen},
  P.~E., \& {Asplund}, M. 2005, \aap, 440, 321

\bibitem[{{Kaufer} {et~al.}(2004){Kaufer}, {Venn}, {Tolstoy}, {Pinte}, \&
  {Kudritzki}}]{kaufer04}
{Kaufer}, A., {Venn}, K.~A., {Tolstoy}, E., {Pinte}, C., \& {Kudritzki}, R.-P.
  2004, \aj, 127, 2723

\bibitem[{{Lacey} \& {Cole}(1993)}]{lacey93}
{Lacey}, C., \& {Cole}, S. 1993, \mnras, 262, 627

\bibitem[{{Lee} {et~al.}(2013){Lee}, {Johnston}, {Tumlinson}, {Sen}, \&
  {Simon}}]{lee13}
{Lee}, D.~M., {Johnston}, K.~V., {Tumlinson}, J., {Sen}, B., \& {Simon}, J.~D.
  2013, \apj, 774, 103

\bibitem[{{Majewski} {et~al.}(2005){Majewski}, {Frinchaboy}, {Kunkel}, {Link},
  {Mu{\~n}oz}, {Ostheimer}, {Palma}, {Patterson}, \& {Geisler}}]{majewski05}
{Majewski}, S.~R. {et~al.} 2005, \aj, 130, 2677

\bibitem[{{Majewski} {et~al.}(2003){Majewski}, {Skrutskie}, {Weinberg}, \&
  {Ostheimer}}]{majewski03}
{Majewski}, S.~R., {Skrutskie}, M.~F., {Weinberg}, M.~D., \& {Ostheimer}, J.~C.
  2003, \apj, 599, 1082

\bibitem[{{Monaco} {et~al.}(2005){Monaco}, {Bellazzini}, {Bonifacio},
  {Ferraro}, {Marconi}, {Pancino}, {Sbordone}, \& {Zaggia}}]{monaco05}
{Monaco}, L., {Bellazzini}, M., {Bonifacio}, P., {Ferraro}, F.~R., {Marconi},
  G., {Pancino}, E., {Sbordone}, L., \& {Zaggia}, S. 2005, \aap, 441, 141

\bibitem[{{Newberg} {et~al.}(2002){Newberg}, {Yanny}, {Rockosi}, {Grebel},
  {Rix}, {Brinkmann}, {Csabai}, {Hennessy}, {Hindsley}, {Ibata}, {Ivezi{\'c}},
  {Lamb}, {Nash}, {Odenkirchen}, {Rave}, {Schneider}, {Smith}, {Stolte}, \&
  {York}}]{newberg02}
{Newberg}, H.~J. {et~al.} 2002, \apj, 569, 245

\bibitem[{{Nissen} \& {Schuster}(1997)}]{nissen97}
{Nissen}, P.~E., \& {Schuster}, W.~J. 1997, \aap, 326, 751

\bibitem[{{Pompeia} {et~al.}(2006){Pompeia}, {Hill}, {Spite}, {Cole}, {Primas},
  {Romaniello}, {Pasquini}, {Cioni}, \& {Smecker-Hane}}]{pompeia06}
{Pompeia}, L. {et~al.} 2006, ArXiv Astrophysics e-prints

\bibitem[{{R. G. Gratton} {et~al.}(2003){R. G. Gratton}, {E. Carretta}, {R.
  Claudi}, {S. Lucatello}, \& {M. Barbieri}}]{gratton03}
{R. G. Gratton}, {E. Carretta}, {R. Claudi}, {S. Lucatello}, \& {M. Barbieri}.
  2003, A\&A, 404, 187

\bibitem[{{Robertson} {et~al.}(2005){Robertson}, {Bullock}, {Font}, {Johnston},
  \& {Hernquist}}]{robertson05}
{Robertson}, B., {Bullock}, J.~S., {Font}, A.~S., {Johnston}, K.~V., \&
  {Hernquist}, L. 2005, \apj, 632, 872

\bibitem[{{Schlaufman} {et~al.}(2009){Schlaufman}, {Rockosi}, {Allende Prieto},
  {Beers}, {Bizyaev}, {Brewington}, {Lee}, {Malanushenko}, {Malanushenko},
  {Oravetz}, {Pan}, {Simmons}, {Snedden}, \& {Yanny}}]{schlaufman09}
{Schlaufman}, K.~C. {et~al.} 2009, \apj, 703, 2177

\bibitem[{Schlaufman {et~al.}(2012)Schlaufman, Rockosi, Lee, Beers, Prieto,
  Rashkov, Madau, \& Bizyaev}]{Schlaufman:2012gh}
Schlaufman, K.~C., Rockosi, C.~M., Lee, Y.~S., Beers, T.~C., Prieto, C.~A.,
  Rashkov, V., Madau, P., \& Bizyaev, D. 2012, Astrophysical Journal, 749, 77

\bibitem[{{Searle} \& {Zinn}(1978)}]{searle78}
{Searle}, L., \& {Zinn}, R. 1978, \apj, 225, 357

\bibitem[{{Sharma} {et~al.}(2010){Sharma}, {Johnston}, {Majewski}, {Mu{\~n}oz},
  {Carlberg}, \& {Bullock}}]{sharma10}
{Sharma}, S., {Johnston}, K.~V., {Majewski}, S.~R., {Mu{\~n}oz}, R.~R.,
  {Carlberg}, J.~K., \& {Bullock}, J. 2010, \apj, 722, 750

\bibitem[{{Shetrone} {et~al.}(2003){Shetrone}, {Venn}, {Tolstoy}, {Primas},
  {Hill}, \& {Kaufer}}]{shetrone03}
{Shetrone}, M., {Venn}, K.~A., {Tolstoy}, E., {Primas}, F., {Hill}, V., \&
  {Kaufer}, A. 2003, \aj, 125, 684

\bibitem[{{Shetrone} {et~al.}(2001){Shetrone}, {C{\^o}t{\'e}}, \&
  {Sargent}}]{shetrone01}
{Shetrone}, M.~D., {C{\^o}t{\'e}}, P., \& {Sargent}, W.~L.~W. 2001, \apj, 548,
  592

\bibitem[{{Smecker-Hane} \& {McWilliam}(2002)}]{smecker-hane02}
{Smecker-Hane}, T.~A., \& {McWilliam}, A. 2002, ArXiv Astrophysics e-prints

\bibitem[{{Somerville} \& {Kolatt}(1999)}]{somerville99}
{Somerville}, R.~S., \& {Kolatt}, T.~S. 1999, \mnras, 305, 1

\bibitem[{{Stephens} \& {Boesgaard}(2002)}]{stephens02}
{Stephens}, A., \& {Boesgaard}, A.~M. 2002, \aj, 123, 1647

\bibitem[{{Tautvai{\v s}ien{\.e}} {et~al.}(2007){Tautvai{\v s}ien{\.e}},
  {Geisler}, {Wallerstein}, {Borissova}, {Bizyaev}, {Pagel}, {Charbonnel}, \&
  {Smith}}]{tautvaisiene07}
{Tautvai{\v s}ien{\.e}}, G., {Geisler}, D., {Wallerstein}, G., {Borissova}, J.,
  {Bizyaev}, D., {Pagel}, B.~E.~J., {Charbonnel}, C., \& {Smith}, V. 2007, \aj,
  134, 2318

\bibitem[{{Ting} {et~al.}(2012){Ting}, {Freeman}, {Kobayashi}, {De Silva}, \&
  {Bland-Hawthorn}}]{ting12}
{Ting}, Y.-S., {Freeman}, K.~C., {Kobayashi}, C., {De Silva}, G.~M., \&
  {Bland-Hawthorn}, J. 2012, \mnras, 421, 1231

\bibitem[{{Unavane} {et~al.}(1996){Unavane}, {Wyse}, \& {Gilmore}}]{unavane96}
{Unavane}, M., {Wyse}, R.~F.~G., \& {Gilmore}, G. 1996, \mnras, 278, 727

\bibitem[{{Venn} {et~al.}(2001){Venn}, {Lennon}, {Kaufer}, {McCarthy},
  {Przybilla}, {Kudritzki}, {Lemke}, {Skillman}, \& {Smartt}}]{venn01}
{Venn}, K.~A. {et~al.} 2001, \apj, 547, 765

\bibitem[{{Venn} {et~al.}(2003){Venn}, {Tolstoy}, {Kaufer}, {Skillman},
  {Clarkson}, {Smartt}, {Lennon}, \& {Kudritzki}}]{venn03}
{Venn}, K.~A., {Tolstoy}, E., {Kaufer}, A., {Skillman}, E.~D., {Clarkson},
  S.~M., {Smartt}, S.~J., {Lennon}, D.~J., \& {Kudritzki}, R.~P. 2003, \aj,
  126, 1326

\bibitem[{{York} {et~al.}(2000){York}, {Adelman}, {Anderson}, {Anderson},
  {Annis}, {Bahcall}, {Bakken}, {Barkhouser}, {Bastian}, {Berman}, {Boroski},
  {Bracker}, {Briegel}, {Briggs}, {Brinkmann}, {Brunner}, {Burles}, {Carey},
  {Carr}, {Castander}, {Chen}, {Colestock}, {Connolly}, {Crocker}, {Csabai},
  {Czarapata}, {Davis}, {Doi}, {Dombeck}, {Eisenstein}, {Ellman}, {Elms},
  {Evans}, {Fan}, {Federwitz}, {Fiscelli}, {Friedman}, {Frieman}, {Fukugita},
  {Gillespie}, {Gunn}, {Gurbani}, {de Haas}, {Haldeman}, {Harris}, {Hayes},
  {Heckman}, {Hennessy}, {Hindsley}, {Holm}, {Holmgren}, {Huang}, {Hull},
  {Husby}, {Ichikawa}, {Ichikawa}, {Ivezi{\'c}}, {Kent}, {Kim}, {Kinney},
  {Klaene}, {Kleinman}, {Kleinman}, {Knapp}, {Korienek}, {Kron}, {Kunszt},
  {Lamb}, {Lee}, {Leger}, {Limmongkol}, {Lindenmeyer}, {Long}, {Loomis},
  {Loveday}, {Lucinio}, {Lupton}, {MacKinnon}, {Mannery}, {Mantsch}, {Margon},
  {McGehee}, {McKay}, {Meiksin}, {Merelli}, {Monet}, {Munn}, {Narayanan},
  {Nash}, {Neilsen}, {Neswold}, {Newberg}, {Nichol}, {Nicinski}, {Nonino},
  {Okada}, {Okamura}, {Ostriker}, {Owen}, {Pauls}, {Peoples}, {Peterson},
  {Petravick}, {Pier}, {Pope}, {Pordes}, {Prosapio}, {Rechenmacher}, {Quinn},
  {Richards}, {Richmond}, {Rivetta}, {Rockosi}, {Ruthmansdorfer}, {Sandford},
  {Schlegel}, {Schneider}, {Sekiguchi}, {Sergey}, {Shimasaku}, {Siegmund},
  {Smee}, {Smith}, {Snedden}, {Stone}, {Stoughton}, {Strauss}, {Stubbs},
  {SubbaRao}, {Szalay}, {Szapudi}, {Szokoly}, {Thakar}, {Tremonti}, {Tucker},
  {Uomoto}, {Vanden Berk}, {Vogeley}, {Waddell}, {Wang}, {Watanabe},
  {Weinberg}, {Yanny}, {Yasuda}, \& {SDSS Collaboration}}]{york00}
{York}, D.~G. {et~al.} 2000, \aj, 120, 1579

\end{thebibliography}

\appendix

\section{The Expectation-Maximization Algorithm}\label{ahp_app}

\subsection{Expectation step}\label{ahp_app:A1}
To implement the algorithm, we first need to derive the expression for the complete data log likelihood, given by Eqn.~\ref{ahp_eq:compLL}, which is conditioned on the data. To do this, it is necessary to decide on a mode of usage for $z_{ij}$. The use of $z$ casts the EM algorithm as either {\it hard} when its value discretely indicates the $f_{j}(x_{i},y_{i})$ of origin or {\it soft} when its value probabilistically indicate the origin of point ($x_{i},y_{i}$) across all $f_{j}$. For this application, we chose to implement a {\it hard} EM algorithm for estimation of $A_{MLE}$ in which $z_{ij}$ has an true value equal to  {\bf 1} if the data point ($x_{i},y_{i}$) comes from model $f_{j}$ or {\bf 0}, otherwise. Thus our overall expectation is
\begin{equation}\label{ahp_eq:compE}
E_{{\bf A}}\Big[\ell({\bf A})|{\bf x,y}\Big]= \sum_{i=1}^{n}\sum_{j=1}^{m}E_{{\bf A}}[z_{ij}|{x_{i},y_{i}}]\,\{{\rm log}\,A_{j} + {\rm log}\,f_{j}(x_{i},y_{i})\}
\end{equation}
where 
\begin{equation}\label{ahp_eq:Ez}
E_{{\bf A}}[z_{ij}|{x_{i},y_{i}}] = \frac{A_{j}\,f_{j}(x_{i},y_{i})}{\sum_{k=1}^{m}A_{k}\,f_{k}(x_{i},y_{i})}
\end{equation}
as defined by Eqn.~\ref{ahp_eq:mixmod}.
Since we are ultimately maximizing Eqn.~\ref{ahp_eq:compE}, the non-constant term, Eqn.~\ref{ahp_eq:Ez}, becomes the component of interest. To iteratively evaluate this expectation, we let $w_{ij}^{(t)}$ be Eqn.~\ref{ahp_eq:Ez} at the $t^{th}$ step:
\begin{equation*}
w_{ij}^{(t+1)} = \begin{dcases}
    \frac{A_{j}\,f_{j}(x_{i},y_{i})}{\sum_{k=1}^{m}A_{k}\,f_{k}(x_{i},y_{i})} & j = 1,\dotsc,m \\  \\
    1 - w_{i1} - \dots -w_{i,m-1} & j = m.
  \end{dcases}
\end{equation*}  

Since ${\bf A}$ is not defined for the first evaluation, we use a random initialization to generate ${\bf w}_{j}^{(0)}$. Here, it should be noted that convergence is not sensitive to the choice of values in our case, though it can be in cases where the likelihood is riddled with local maxima. If we examine the expression above, we can conceptually define the mechanism for maximization as a ``ratcheting up'' of $E_{A}[z_{ij}|{x_{i},y_{i}}]$ values by maximizing $A_{j}\,f_{j}(x_{i},y_{i})$ with respect to $\sum_{k=1}^{m}A_{k}\,f_{k}(x_{i},y_{i})$. Derivation of the maximization expression is discussed below.

\subsection{Maximization step}\label{ahp_app:A2}
Above we defined an explicit formulation for the expected log-likelihood (Eqn.~\ref{ahp_eq:Ez}) given a single parameter ${\bf A}$ and the data ({\bf x,y}). The argument of the maximum of Eqn.~\ref{ahp_eq:Ez} at each iteration $t$ provides an estimate that approaches the MLE of ${\bf A}$, and is given by:
\begin{equation}\label{ahp_eq:argmax}
{\bf A}^{(t)} = \underset{{\bf A}}{\operatorname{argmax}}\Big[\ell({\bf A})|{\bf x,y,}{\bf A}^{(t-1)}\Big].
\end{equation}

Accounting for the $m$-1 free parameters of ${\bf A}$, differentiation of Eqn.~\ref{ahp_eq:compE} with Eqn.~\ref{ahp_eq:Ez} proceeds, for $k = 1,\dotsc,m-1$, as:
\begin{equation*}
\frac{\partial}{\partial A_{k}}\,E_{A}\Big[\ell({\bf A})|{\bf x,y}\Big]= \sum_{i=1}^{n}\Big\{w_{ik}^{(t-1)}\frac{1}{A_{k}} -w_{im}^{(t-1)}\frac{1}{1-A_{1}-\dots-A_{m-1}}\Big\}
\end{equation*}
where the first term in the summation accounts all values of $k\leq m$ and the second term eliminates over-counting of the $1^{st}$-term at $k = m$. The derivative of an argmax is always equal to zero since we are taking a derivative at the maximum point of the function in question (in our case the expectation of the log-likelihood). Thus, we can expand the summation of data points and equate the terms described above to one another $$\frac{1}{A_{k}}\sum_{i=1}^{n}w_{ik}^{(t-1)}=\frac{1}{1-A_{1}-\dots-A_{m-1}}\sum_{i=1}^{n}w_{im}^{(t-1)}.$$
Consequently, these terms being equal means that every $k\leq m$ term is equal to each as shown below $$\frac{1}{A_{k}}\sum_{i=1}^{n}w_{ik}^{(t-1)}=\dots=\frac{1}{A_{m-1}}\sum_{i=1}^{n}w_{i,m-1}^{(t-1)}=c$$ and $${\bf A}_{k}^{(t)} = \frac{\sum_{i=1}^{n}w_{ik}^{(t-1)}}{c}$$ where $c$ is some constant.

The unknown constant $c$ appears problematic, but, because $\sum_{j=1}^{m}A_{j} = 1$, algebraic manipulation reveals that $c = n$, yielding a final solution that can be numerically evaluated:
\begin{align}
{\bf A}_{k}^{(t)} &= \frac{\sum_{i=1}^{n}w_{ik}^{(t-1)}}{n}\label{ahp_eq:diffE1} \\
{\bf A}_{m}^{(t)} &= 1-A_{1}-\dots-A_{m-1}.\label{ahp_eq:diffE2}
\end{align}

Finally, to implement this algorithm, we simply compute an initial value for ${\bf A}$, inserting each component, ${A}_{j}$, into an $w_{ik}^{t}$ equal to Eqn.~\ref{ahp_eq:Ez} (i.e. with $k$ initially identical to $j$) and then compute that expression with Eqn.~\ref{ahp_eq:diffE1} to calculate each new corresponding ${A}_{k}$. This process is repeated unto our iteration criterion is met.

In our case, computation of ${\bf A}\rightarrow{\bf A}_{EM}$ converges relatively quickly for all starting values: on the order of 600 iterations, or half a minute, for $n = 1000$ (given our stopping criteria). Large ${A}_{EM,k}$ values typically emerge after two or three iterations, and most change, absolutely speaking, occurs in the first fifty to one hundred iterations. For error estimation, we can provide values for the minimum error possible through an inversion of the Fisher information matrix (see Appendix~\ref{ahp_app:A3} for brief derivation). Although we have an idea of what the best possible errors are, such values exclude the use of more standard approaches to assessments of parameter estimation, like the reduced $\chi^{2}$ statistic. 

\subsection{Derivation of the minimum error on EM estimates}\label{ahp_app:A3}
The asymptotic covariance matrix of $\hat{\bf A}_{EM}$ can be approximated by the inverse of the observed Fisher information matrix, $I$.

As $A_{EM,m} = 1-\sum_{j=1}^{(m-1)}A_{EM,j}$, there are only $m-1$ free parameters. Thus let ${\bf A}_{EM}^{\prime}=(A_{EM,1},\ldots,A_{EM,(m-1)})$. Using $f_{ij} = f_j(x_i,y_i)$ for brevity, the likelihood can then be expressed as:
\begin{eqnarray}\label{llinfo}
	 \ell({\bf A}_{EM}^{\prime}) = \sumn \log \Bigg\{ \Big( \summo A_{EM,j} f_{ij} \Big) + (1-A_{EM,1},\ldots,A_{EM,(m-1)}) f_{im} \Bigg\} 
\end{eqnarray}

The observed information matrix, $I$, is the $(m-1)\times (m-1)$ negative hessian of Eqn.~\ref{llinfo}, evaluated at the observed data points:
\mtx{I({\bf A}_{EM}^{\prime}|\vx,\vy)=-\frac{\partial^2 \ell({\bf A}_{EM}^{\prime})}{\partial {\bf A}_{EM}^{\prime} \partial {\bf A}_{EM}^{\prime T}}}{-}{cccc}{
	\frac{\partial^2 \ell({\bf A}_{EM}^{\prime})}{\partial^2 A_{EM,1}} & \hesslld{1}{2} & \ldots & \hesslld{1}{(m-1)}	\\
	\vdots & \vdots & & \vdots	\\
	\hesslld{(m-1)}{1} & \hesslld{(m-1)}{2} & \ldots & \frac{\partial^2 \ell({\bf A}_{EM}^{\prime})}{\partial^2 A_{EM,(m-1)}}
}

where
\eqn{
	\frac{\partial \ell({\bf A}_{EM}^{\prime})}{\partial A_{EM,k}} = \sumn \frac{f_{ik} - f_{im}}{\summ A_{EM,j} f_{ij}}	\hspace{10mm}\text{and}\hspace{10mm}	\frac{\partial^2 \ell({\bf A}_{EM}^{\prime})}{\partial A_{EM,k} \partial A_{EM,r}} = -\sumn \frac{(f_{ik}-f_{im}) (f_{ir}-f_{im})}{(\summ A_{EM,j} f_{ij})^2}
}
with $1 \le r \le m-1$ such that ($k$,$r$) represents the index of the observed information matrix $I$.

The observed information matrix of ${\bf A}_{EM}^{\prime}$ yields the following estimates for covariance and correlation for all $m$ estimated weights in $\hat{\bf A}_{EM}$:

\eqnset{\text{Cov}(\hat{A}_{EM,p},\hat{A}_{EM,q})}{}{ll}{
	\big[I^{-1}(\hat{\bf A}_{EM}^{\prime}) \big]_{Eq}				& p,q<m	\\
	-\sum\limits_{j=1}^{m-1} \text{Cov}(\hat{A}_{EM,j},\hat{A}_{EM,q})		& p=m,q<m	\\
	\sum\limits_{j=1}^{m-1} \sum\limits_{k=1}^{m-1} \text{Cov}(\hat{A}_{E,j},\hat{A}_{EM,q})		& p, q=m	\\
}

\eqn{
	\text{Var}(\hat{A}_{EM,j}) &= \sigma^2_j = \Bl  \text{Cov}(\hat{\bf A}_{EM}) \Br_{jj}
}

\end{document}